\documentclass[%
 reprint,
 amsmath,amssymb,
 aps,
]{revtex4-2}

\usepackage{graphicx}
\usepackage{bm}
\usepackage{color}

\begin{document}

\preprint{APS/123-QED}

\title{Influence of quantum decoherence on black hole shadow visibility and time domain echoes}

\author{Sen Guo}
\affiliation{College of Physics and Optoelectronic Engineering, Chongqing Normal University, Chongqing 401331, People's Republic of China}

\author{Xin Li}
\affiliation{College of Physics and Optoelectronic Engineering, Chongqing Normal University, Chongqing 401331, People's Republic of China}

\author{Yu Liang}
\affiliation{School of Big Data and Artificial Intelligence, Fuyang University of Technology, Fuyang 236000, People's Republic of China}

\author{Kai Lin}
\affiliation{Universidade Federal de Campina Grande, Campina Grande, PB, Brasil, \\ Instituto de F\'isica, Universidade de S\~ao Paulo, S\~ao Paulo, Brasil}%

\author{Li-Fang Li}
\email{lilifang@imech.ac.cn (Corresponding author)}
\affiliation{Center for Gravitational Wave Experiment, National Microgravity Laboratory, Institute of Mechanics, Chinese Academy of Sciences, Beijing 100190, People's Republic of China}


\date{\today}

\begin{abstract}
The geometric shadow of a black hole cannot observationally distinguish horizonless exotic compact objects. We break this degeneracy by mapping the fundamental quantum decoherence of the horizon (the Danielson-Satishchandran-Wald mechanism) into a macroscopic effective reflectivity $\mathcal{R}$. Utilizing Kerr ray tracing, we show that a partially reflective boundary ($\mathcal{R}>0$) generates delayed electromagnetic echoes under both specular and diffuse (Lambertian) conditions. In the spatial domain, these echoes illuminate the inner shadow and alter low-frequency visibilities, while leaving metric-governed photon subrings unchanged. The signal strength scales with the membrane compactness $\epsilon$ as $\epsilon^{\Gamma}$, where $\Gamma=0$ for specular reflection and $\Gamma\to1$ for diffuse re-emission due to escape-cone collapse. This brings specular reflectivities down to $\mathcal{R}\simeq0.12$ (ngEHT) and $0.04$ (BHEX) within reach of static imaging. For Sagittarius A$^\ast$, we translate these echoes into a robust time-domain signature. Using an Autocorrelation Function (ACF) pipeline, we demonstrate that single-flare echoes can be cleanly extracted from severe accretion red noise, achieving a $3\sigma$ detection for amplitudes $\mathcal{A}\gtrsim0.3$ within 15 to 33 hours of continuous monitoring. This framework provides a directly deployable pathway to test black hole unitarity.
\end{abstract}

\maketitle


\section{Introduction}
\label{sec:intro}
\par
The defining macroscopic feature of a black hole in general relativity is the event horizon. Over the past decade, the Event Horizon Telescope (EHT) collaboration has achieved groundbreaking milestones by resolving \textcolor{black}{horizon-scale} millimeter images of the supermassive black holes M87$^\ast$ and Sagittarius A$^\ast$ (Sgr A$^\ast$)~\cite{EventHorizonTelescope:2019dse,EventHorizonTelescope:2022wkp}. These observations reveal a bright emission ring encircling a central brightness depression, consistent with the theoretical shadow produced by strong gravitational lensing~\cite{EventHorizonTelescope:2026cny}. However, \textcolor{black}{such} an interferometric shadow is fundamentally governed by the unstable photon sphere rather than the event horizon itself. A variety of horizonless exotic compact objects \textcolor{black}{(ECOs, such as boson stars, gravastars, or wormholes)} can cast nearly identical shadows if they are sufficiently compact~\cite{Li:2025awg,Rosa:2024bqv,Aimar:2025uia,He:2025qmq,Vincent:2015xta,Zeng:2025nmu,Chen:2024tss,Huang:2024bbs,Delijski:2022jjj,Wielgus:2020uqz,Peng:2021osd}. Consequently, current spatial imaging confirms the existence of the photon ring but leaves the absolute nature of the true \textcolor{black}{event} horizon observationally unverified. The definitive physical attribute that distinguishes a true black hole from an \textcolor{black}{ECO} is not the geometric size of the shadow, but the boundary condition: whether the surface strictly absorbs or partially reflects~\cite{EventHorizonTelescope:2022xqj}.

\par
The nature of this boundary is intrinsically tied to the black hole information \textcolor{black}{paradox} and the unitarity of quantum mechanics~\cite{Hawking:1976ra,Almheiri:2012rt}. Recently, Danielson \textit{et~al.} demonstrated that the mere presence of a Killing horizon fundamentally decoheres a quantum system held in a spatial superposition outside it~\cite{Danielson:2022tdw,Danielson:2022sga}. In this framework, the continuous emission of soft entangling photons (or gravitons) carries which-path information across the horizon~\cite{Belenchia:2018szb}. Because \textcolor{black}{a true} horizon acts as a perfect \textcolor{black}{and} irreversible information sink, the off-diagonal coherence of the external quantum state decays exponentially~\cite{Gralla:2023oya}. This decoherence effect rigorously establishes that the horizon's capacity to irreversibly absorb information is an inherent property of the spacetime geometry. However, this phenomenon has been widely regarded as fundamentally unobservable, as it acts exclusively on the \textcolor{black}{quantum} state of a \textcolor{black}{single} microscopic system and leaves no macroscopic imprint on the background optical appearance~\cite{Danielson:2022tdw,Danielson:2022sga,Fahn:2026rei}.

\par
In this work, we construct \textcolor{black}{a} quantitative bridge linking this theoretically unobservable quantum decoherence to macroscopically falsifiable black hole imaging~\cite{Vagnozzi:2022moj}. We introduce a phenomenological parameter, \textcolor{black}{namely} the effective horizon reflectivity $\mathcal{R} \in [0,1]$, to establish an exact macroscopic-microscopic correspondence~\cite{Maggio:2017ivp,Cardoso:2019rvt}. \textcolor{black}{Dictated} by energy conservation, the \textcolor{black}{radiation} field interacting with the near-horizon region is strictly partitioned. The absorbed fraction $(1-\mathcal{R})$ acts as the irreversible information sink, driving \textcolor{black}{the} quantum decoherence at a rate proportional to the surface gravity $\kappa$~\cite{Danielson:2022tdw,Gralla:2023oya}. Concurrently, the reflected fraction $\mathcal{R}$ generates a sequence of delayed electromagnetic echoes \textcolor{black}{that return} to the exterior spacetime~\cite{Cardoso:2016rao,Abedi:2016hgu}. This establishes a direct one-to-one mapping: a true black hole ($\mathcal{R}=0$) completely decoheres external quantum states and dictates a completely dark inner shadow~\cite{Chael:2021rjo}. Conversely, a horizonless membrane ($\mathcal{R}>0$) suppresses continuous decoherence while illuminating the inner shadow via geometrically delayed echoes~\cite{Abedi:2018npz}. Therefore, the theoretically unobservable quantum decoherence rate can be directly constrained by classical imaging and time-domain observables~\cite{Wielgus:2022heh}.

\par
Motivated by this theoretical mapping, the primary goal of this paper is to translate these discrete temporal echoes into steady state spatial \textcolor{black}{interferometric signatures} and dynamic light curves, thereby evaluating their detectability. Utilizing \textcolor{black}{an adaptive} general relativistic ray tracing \textcolor{black}{(GRRT) algorithm} in the Kerr spacetime, we map the corresponding signatures into the parameter space of current and next-generation Very Long Baseline Interferometry (VLBI) arrays~\cite{Moscibrodzka:2017lcu,Chael:2022meh,Roelofs:2022lux}. We demonstrate that the boundary reflectivity uniquely elevates the inner shadow filling and modulates the low-frequency visibility envelope, while preserving the \textcolor{black}{metric-governed, high-frequency} photon ring signatures~\cite{Johnson:2019ljv}. Furthermore, we define a unified horizon discovery space spanned by the reflectivity and the boundary compactness~\cite{Maggio:2017ivp,Cardoso:2016rao}. Our analysis reveals a strict observational complementarity dictated by the physical timescales of the targets: while the multi day echo delays of M87$^\ast$ require ultra deep static imaging to probe the inner shadow, the echoes of Sgr A$^\ast$ manifest on minute timescales. This temporal compression transforms Sgr A$^\ast$ into an ideal time domain laboratory, where high cadence monitoring can capture multiple reflection cycles~\cite{Abuter:2018uum}. Together, these strategies aim to provide quantitative observational limits on the existence of an information-absorbing horizon.

\par
The paper is organized as follows. Section~\ref{sec:2} establishes the theoretical framework, detailing the highly adaptive ray-tracing baseline in the Kerr spacetime and the formal derivation of the quantum classical dictionary. Section~\ref{sec:3} translates the time-domain echoes into static spatial interferometric features, isolating the inner shadow filling from the universal photon subrings, and defines the horizon discovery space for future VLBI arrays. Section~\ref{sec:4} extends the framework to \textcolor{black}{a} time-domain detection pipeline, demonstrating the feasibility of extracting \textcolor{black}{periodic} Sgr A$^\ast$ reflection echoes \textcolor{black}{from stochastic accretion red noise} using autocorrelation \textcolor{black}{function (ACF)} techniques\textcolor{black}{, highlighting the detectability thresholds under realistic multi-wavelength observational conditions}. Finally, we summarize our main conclusions in Section~\ref{sec:5}.

\section{Mapping Quantum Decoherence to Macroscopic Geodesics}
\label{sec:2}
\par
This section establishes the theoretical framework linking the macroscopic observables of black hole imaging to the microscopic quantum properties of the boundary. \textcolor{black}{Within} the Danielson-Satishchandran-Wald (DSW) framework, a perfect Killing horizon irreversibly acts as an information sink, continuously decohering external quantum superpositions at a rate $\Gamma_{\rm dec}^{(0)} \propto \kappa$, where $\kappa$ is the surface gravity~\cite{Danielson:2022tdw,Danielson:2022sga}. To macroscopically parameterize this boundary condition, we introduce a single effective reflectivity $\mathcal{R} \in [0, 1]$ governing the near-horizon radiation field. \textcolor{black}{Dictated} by energy conservation, the incident radiation flux $\mathcal{F}_{\rm in}$ is strictly partitioned into an absorbed fraction $\mathcal{F}_{\rm abs} = (1-\mathcal{R})\mathcal{F}_{\rm in}$ and a reflected fraction $\mathcal{F}_{\rm ref} = \mathcal{R}\mathcal{F}_{\rm in}$~\cite{Maggio:2017ivp,Cardoso:2019rvt}. Consequently, the physical decoherence rate is modulated to $\Gamma_{\rm dec} = (1-\mathcal{R})\Gamma_{\rm dec}^{(0)}$. This yields a rigorous quantum-classical dichotomy: the true black hole limit ($\mathcal{R} = 0$) implies absolute information loss and maximal decoherence, dictating a completely dark inner shadow with no delayed echoes~\cite{Chael:2021rjo}. Conversely, a horizonless compact object ($\mathcal{R} > 0$) suppresses continuous decoherence while reflecting electromagnetic information back to the exterior universe after a geometric time delay, thus macroscopically illuminating the inner shadow~\cite{Abedi:2018npz}.

\par
To quantitatively trace these returning macroscopic echoes, we must rigorously model the null geodesic flow in the background spacetime. We adopt geometric units $G = c = M = 1$ and utilize Boyer-Lindquist coordinates $(t, r, \theta, \phi)$ with the metric signature $(-, +, +, +)$. The covariant components of the Kerr metric $g_{\mu\nu}$ are given by
\begin{eqnarray}
g_{tt} &=& -\left(1 - \frac{2r}{\Sigma}\right), \quad g_{rr} = \frac{\Sigma}{\Delta}, \quad g_{t\phi} = -\frac{2ar\sin^2\theta}{\Sigma}, \nonumber \\
g_{\phi\phi} &=& \left(r^2 + a^2 + \frac{2a^2 r \sin^2\theta}{\Sigma}\right)\sin^2\theta, \quad g_{\theta\theta} = \Sigma,
\label{eq:1}
\end{eqnarray}
where $\Sigma \equiv r^2 + a^2\cos^2\theta$, $\Delta \equiv r^2 - 2r + a^2$, and $a$ is the dimensionless spin parameter. The propagation of high-frequency electromagnetic radiation is governed by the null geodesic Hamiltonian $\mathcal{H} = \frac{1}{2}g^{\mu\nu}p_\mu p_\nu = 0$, where $p_\mu$ is the photon four-momentum.

\par
\begin{figure*}[htbp]
\centering
\includegraphics[width=15cm,height=4.5cm]{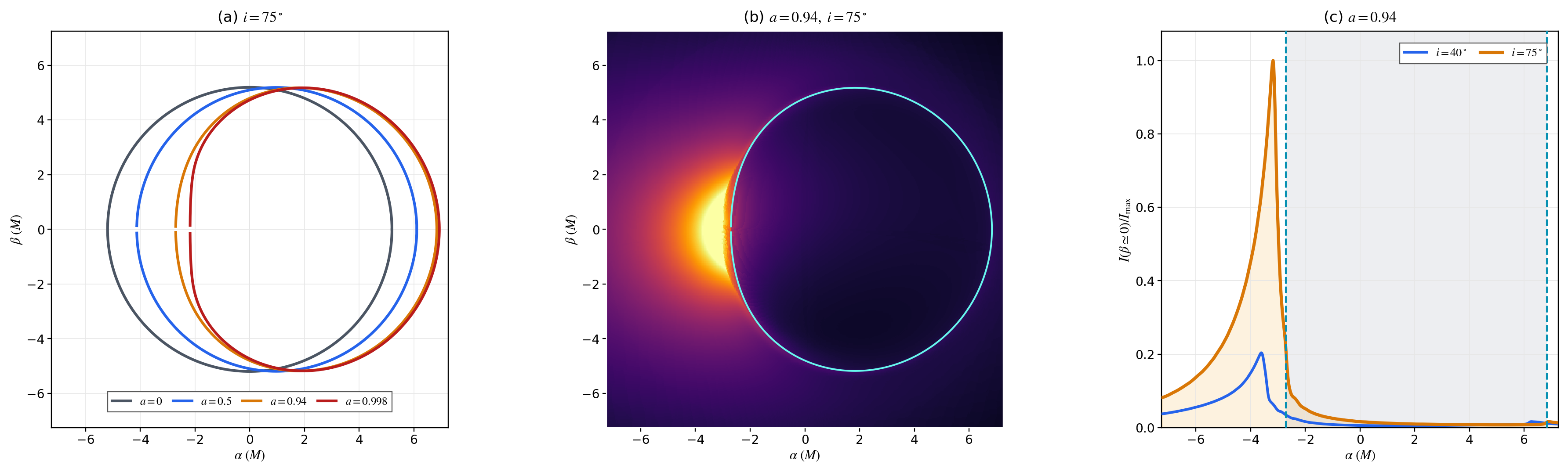}
\caption{Geometric validation of the \textcolor{black}{custom} GRRT baseline in the Kerr spacetime. (a) The mathematically exact critical curves $(\alpha, \beta)$ characterizing the theoretical shadow boundary for varying spin parameters $a$ at an observer inclination of $i=75^\circ$. (b) \textcolor{black}{The synthesized} GRRT image for a standard black hole ($a=0.94, i=75^\circ$) illuminated by a geometrically thick \textcolor{black}{and} optically thin accretion model (emissivity $\propto r^{-3}e^{-\cos^2\theta/2H^2}$ with $H=0.5$ for $r_{\rm ISCO}\le r\le12\,GM/c^2$, emitters on Keplerian orbits), with the analytic critical curve (cyan solid line) flawlessly superimposed. (c) The horizontal intensity cross-section at $\beta \simeq 0$. The sharp exponential decay accurately aligns with the theoretical shadow edge (vertical dashed lines), confirming the numerical fidelity of the geodesic solver in resolving the extreme lensing regime before \textcolor{black}{any} boundary modifications are introduced.}
\label{fig:1}
\end{figure*}

The isometries of the Kerr spacetime, generated by the Killing vector fields $\partial_t$ and $\partial_\phi$, alongside the hidden symmetry associated with the Killing tensor, guarantee the complete separability of the Hamilton-Jacobi equation. This \textcolor{black}{yields} three conserved quantities along any photon trajectory: the energy $E = -p_t$, the axial angular momentum $L = p_\phi$, and the Carter constant $Q$. Defining the energy-rescaled impact parameters $\xi \equiv L/E$ and $\eta \equiv Q/E^2$, the theoretical boundary of the black hole shadow is exclusively determined by the unstable spherical photon orbits~\cite{Gralla:2019drh}. These orbits occur at a constant Boyer-Lindquist radius $r = r_p$, satisfying the radial potential conditions $R(r_p) = 0$ and $R'(r_p) = 0$\textcolor{black}{, where $R(r)=[r^2+a^2-a\xi]^2-\Delta[\eta+(\xi-a)^2]$ is the radial potential in units of $E^2$}. Solving these conditions yields the critical impact parameters $(\xi_c, \eta_c)$ parameterized by the photon sphere radius $r_p$:
\begin{eqnarray}
&&\xi_c(r_p) = -\frac{r_p^3 - 3r_p^2 + a^2r_p + a^2}{a(r_p-1)}, \nonumber \\
&&\eta_c(r_p) = -\frac{r_p^3(r_p^3 - 6r_p^2 + 9r_p - 4a^2)}{a^2(r_p-1)^2}.
\label{eq:2}
\end{eqnarray}
\textcolor{black}{Projected} onto the celestial plane of a distant observer situated at an inclination angle $i$, the Cartesian Bardeen coordinates $(\alpha, \beta)$ of this critical curve are given by $\alpha = -\xi_c/\sin i$ and $\beta = \pm\sqrt{\eta_c + a^2\cos^2 i - \xi_c^2\cot^2 i}$. This critical curve establishes a fundamental geometric separatrix. Photons originating from the accretion flow with impact parameters outside this closed curve escape to infinity, forming the extended image~\cite{Ouyang:2025qru}. In contrast, trajectories mapped inside this curve are destined to intersect the \textcolor{black}{event} horizon. Therefore, the observational signatures of the quantum boundary condition $\mathcal{R}$ are strictly confined within this theoretical boundary, i.e., the inner shadow~\cite{Zhen:2025nah}. As demonstrated in Fig.~\ref{fig:1}, the numerical shadow boundary synthesized by our \textcolor{black}{custom} GRRT code exhibits remarkable quantitative agreement with the analytic critical curve. The intensity cross-section (Fig.~\ref{fig:1}c) confirms the mathematically pblackicted sharp truncation at the theoretical shadow edge, validating the precision of our underlying geodesic solver in resolving the extreme gravitational lensing near $r_p$, before \textcolor{black}{any} boundary modifications are introduced.

\par
Having established the critical curve as the geometric separatrix, we introduce a stationary, axisymmetric reflective membrane positioned infinitesimally outside the event horizon at $r_b = r_+ (1+\epsilon)$, where $r_+ = 1+\sqrt{1-a^2}$ is the outer horizon radius and $\epsilon \ll 1$ represents the dimensionless boundary compactness. Photons striking this membrane undergo local specular reflection ($p_r \to -p_r$, while conserving $p_t, p_\theta, p_\phi$) and suffer an intensity attenuation strictly scaled by the effective reflectivity $\mathcal{R}$. \textcolor{black}{Because $g^{r\mu}=0$ for $\mu\neq r$ in Boyer--Lindquist coordinates, this map is elastic in every stationary frame and conserves $E$, $L$ and $Q$. Numerically, the outgoing radial momentum is recomputed at $r_b$ from the null condition, $p_r=+E\sqrt{R(r_b)}/\Delta(r_b)$, so that the reflected ray remains exactly null. Specular reflection is one limiting case of the unknown microscopic reflection law; in Sec.~\ref{sec:3} we also consider the opposite limit of diffuse (Lambertian) re-emission.} \textcolor{black}{It is worth noting that since the wavelength of the soft modes in the relevant infrablack window significantly exceeds the typical microscopic scale of the membrane, the reflection is largely dominated by the macroscopic effective barrier. This macroscopic dominance makes the frequency dependence negligible and robustly justifies the gray-body approximation ($\mathcal{R}(\omega) \approx \mathcal{R}$) adopted here.}

\par
\begin{figure*}[htbp]
\centering
\includegraphics[width=15cm,height=4.5cm]{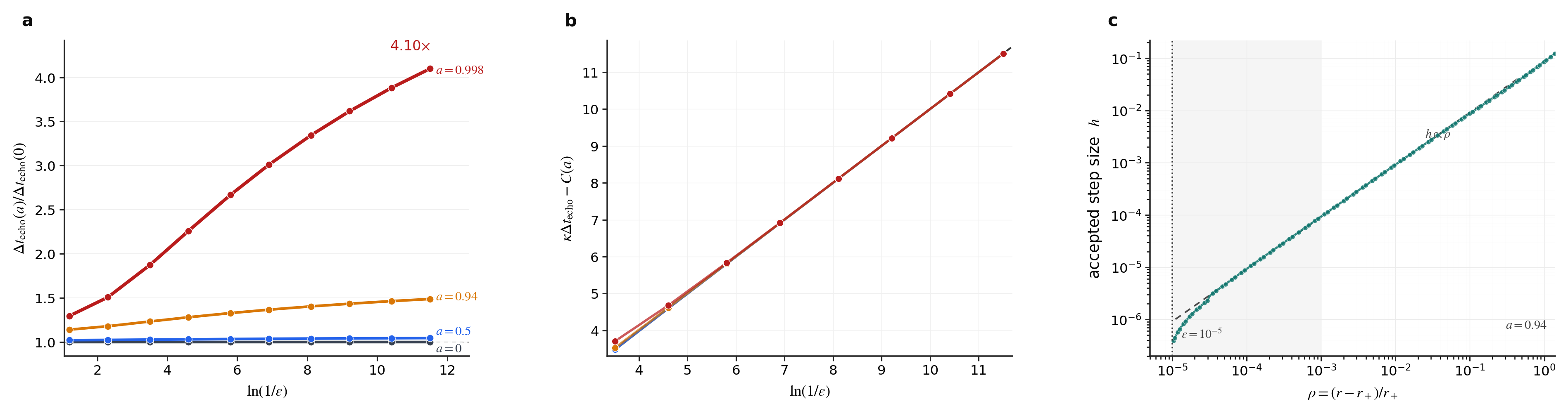}
\caption{Numerical verification of the near-horizon \textcolor{black}{echo-delay} physics and time-dilation universality. \textcolor{black}{(a) Ratio of the ray-traced echo delay to its Schwarzschild counterpart, $\Delta t_{\rm echo}(a)/\Delta t_{\rm echo}(0)$, versus $\ln(1/\epsilon)$.} Near-extremal spins ($a=0.998$) radically amplify the delay due to the vanishing surface gravity ($\kappa \to 0$)\textcolor{black}{, asymptotically growing toward $\kappa(0)/\kappa(a)$}. (b) Normalizing the echo delay by $\kappa$ universally collapses the slopes for all spin parameters ($\kappa \Delta t_{\rm echo} \propto \ln(1/\epsilon)$), directly verifying the theoretical scaling. (c) Trajectory of the accepted affine step size $h$ dynamically evaluated by the DP45 integrator \textcolor{black}{(with the null constraint rigorously re-imposed)}. The step size strictly collapses proportionally to the fractional distance $\rho = (r-r_+)/r_+$, successfully resolving the analytical stiffness \textcolor{black}{and avoiding massive local truncation errors}.}
\label{fig:2}
\end{figure*}

Extracting these escaping echoes requires numerically integrating the null geodesic flow exceedingly deep into the gravitational potential well (down to $\epsilon \sim 10^{-5}$). However, this extreme regime intrinsically introduces severe numerical stiffness. Governed by Hamilton's equations $\dot{p}_r = -\partial \mathcal{H} / \partial r = -\frac{1}{2}(\partial_r g^{\mu\nu})p_\mu p_\nu$, the radial momentum derivative near the horizon is \textcolor{black}{dominated by the terms that diverge at the horizon, i.e., the $t$--$\phi$ block of the inverse metric and, on the null shell, the $g^{rr}p_r^2$ term, which contributes equally}. Utilizing the asymptotic expansion of the inverse metric near $r_+$, the local acceleration diverges as:
\begin{equation}
\dot{p}_r \simeq -\frac{E^2 (r_+^2+a^2-a\xi)^2}{\Sigma_+} \frac{\Delta'(r)}{\Delta^2(r)}.
\label{eq:3}
\end{equation}
Because $\Delta(r_+) = 0$ while its derivative $\Delta'(r_+) = 2(r_+ - 1) \neq 0$, the spatial gradient mathematically diverges as $\partial_r g^{tt} \propto (r - r_+)^{-2} \to \infty$. \textcolor{black}{Consequently, standard fixed-step integrators---including traditional symplectic algorithms---fail in this regime. While traditional symplectic integrators excel in long-term Hamiltonian conservation for stable orbital dynamics, they inevitably fail in this highly localized plunging regime due to massive local truncation errors originating from the divergent spatial gradients, and the consequent violations of the null constraint $\mathcal{H} = 0$. To rigorously resolve this, we employ the Dormand-Prince (DP45) adaptive Runge-Kutta algorithm. Therefore, the adaptive step-size mechanism is mathematically indispensable here to strictly preserve the null constraint}: by concurrently evaluating a 5th-order solution alongside a 4th-order embedded estimate, the local error norm $\|\hat{e}\|$ dynamically regulates the affine step size $h$:
\begin{equation}
h_{\rm new} = {\rm clip}\left[ h\cdot{\rm clip}\left(0.9 \|\hat{e}\|^{-1/5}, 0.2, 5\right), h_{\rm min}, h_{\rm max} \right].
\label{eq:4}
\end{equation}
As demonstrated in Fig.~\ref{fig:2}(c), this adaptive scheme is a mathematical necessity: as the photon plunges toward $r_b$, the accepted step size $h$ dynamically collapses by orders of magnitude, $h\propto\rho$. \textcolor{black}{To suppress truncation errors that otherwise severely violate the null constraint at small $\epsilon$, we strictly project $p_r$ onto $\mathcal{H}=0$ after each step. This bounds the fractional error below $10^{-15}$, yielding echo delays consistent with direct $dt/dr$ quadrature to better than $10^{-3}$.}

\par
\textcolor{black}{The images of Sec.~\ref{sec:3} are computed using an independent formulation in Mino time $\sigma$~\cite{Mino:2003yg}. Because the separated radial and angular potentials, $R(r)$ and $\Theta(\cos\theta)$, remain polynomial and regular at $r_+$, this approach fundamentally circumvents the near-horizon stiffness of Eq.~(\ref{eq:3}). We integrate the second-order geodesic equations with the DP45 scheme, re-impose the null constraint after each step, and cleanly restart the reflected rays with $dr/d\sigma=+\sqrt{R(r_b)}$. Every captublack ray reflects exactly once and escapes. For $\mathcal{R}=0$, this Mino-time solver perfectly matches our Boyer--Lindquist integrator to a median pixel difference of $2\times10^{-4}$ (and $10^{-5}$ in total flux).}

\par
The fundamental macroscopic observable associated with this microscopic boundary $\epsilon$ is the geometric time delay of the echoes, $\Delta t_{\rm echo}$. Along the principal null congruences, the coordinate time propagation is governed by the Kerr tortoise coordinate $r_* = \int \frac{r^2+a^2}{\Delta} dr$. Applying partial fraction decomposition to $\Delta = (r-r_+)(r-r_-)$ yields the exact radial integral:
\begin{equation}
r_*(r) = r + \frac{r_+^2+a^2}{r_+ - r_-}\ln(r-r_+) - \frac{r_-^2+a^2}{r_+ - r_-}\ln(r-r_-).
\label{eq:5}
\end{equation}
Recognizing the exact thermodynamic definition of the Kerr surface gravity, $\kappa = \frac{r_+ - r_-}{2(r_+^2+a^2)}$, the coefficient of the dominant near-horizon logarithmic singularity identically simplifies to $\frac{1}{2\kappa}$. For a macroscopic round-trip extending from the external emission region down to the membrane $r_b = r_+(1+\epsilon)$ and back, the temporal delay is \textcolor{black}{overwhelmingly} dominated by this asymptotic regime. Thus, the echo delay fundamentally scales with the intrinsic spacetime parameters as:
\begin{equation}
\Delta t_{\rm echo} \simeq 2 \left| \int_{r_{\rm start}}^{r_b} dr_* \right| \simeq \frac{1}{\kappa}\ln\left(\frac{1}{\epsilon}\right) + \mathcal{C}(a, \theta).
\label{eq:6}
\end{equation}
Figure~\ref{fig:2}(a) and (b) provide strict numerical verification of the analytical relation in Eq.~(\ref{eq:6}). The integrated delay $\Delta t_{\rm echo}$ dynamically synthesized by the GRRT exhibits a pristine linear dependence on $\ln(1/\epsilon)$. Crucially, upon multiplying by the analytical surface gravity $\kappa$, the disparate slopes across different spin parameters perfectly collapse into a universal unity slope (Fig.~\ref{fig:2}b). This definitively proves that our adaptive geodesic solver seamlessly captures the exact general relativistic time-dilation physics \textcolor{black}{deep into} the \textcolor{black}{near-}horizon limit.

\par
Having established the strictly relativistic kinematic baseline for the echo delay ($\Delta t_{\rm echo}$), we now introduce the boundary dynamics ($\mathcal{R}$) to construct the formal dictionary connecting macroscopic imaging to the microscopic quantum properties of the horizon. In the context of the black hole information paradox, if a spatial quantum superposition of size $d$ is maintained outside the black hole for a proper time $T$, the horizon-induced decoherence rate $\Gamma_{\rm dec}$ is fundamentally driven by the emission and irreversible absorption of entangling soft radiation~\cite{Danielson:2022tdw}. For a true black hole ($\mathcal{R}=0$), this rate is strictly proportional to the surface gravity $\kappa$~\cite{Gralla:2023oya}. However, when the boundary is partially reflective ($\mathcal{R}>0$), only the transmitted fraction $(1-\mathcal{R})$ of this radiation acts as a continuous information sink~\cite{Danielson:2022sga}. Consequently, the effective quantum decoherence rate is rigorously bottlenecked by the classical boundary absorptivity:
\begin{equation}
\frac{\Gamma_{\rm dec}}{\kappa} \propto 1 - \mathcal{R}.
\label{eq:7}
\end{equation}

\textcolor{black}{We emphasize that Eq.~(\ref{eq:7}) represents a phenomenological correspondence rather than a derivation from a UV-complete effective field theory. In open quantum systems, environmental decoherence fundamentally arises from tracing out degrees of freedom into a causal sink. Dictated by probability conservation, the reflected fraction $\mathcal{R}$ of incident modes retains its information in the exterior, while only the absorbed fraction $(1-\mathcal{R})$ is permanently traced out, directly bottlenecking the decoherence rate.}

\textcolor{black}{Crucially, this analogy applies specifically within the extreme soft infrablack window ($\omega \lesssim \kappa$) of the DSW mechanism. Because these low-energy wavelengths vastly exceed any fundamental microscopic boundary scale, the interaction is overwhelmingly governed by macroscopic energy partitioning. This robustly justifies the frequency-independent gray-body approximation ($\mathcal{R}(\omega) \approx \mathcal{R}$).}

\par
\begin{figure*}[htbp]
\centering
\includegraphics[width=15cm]{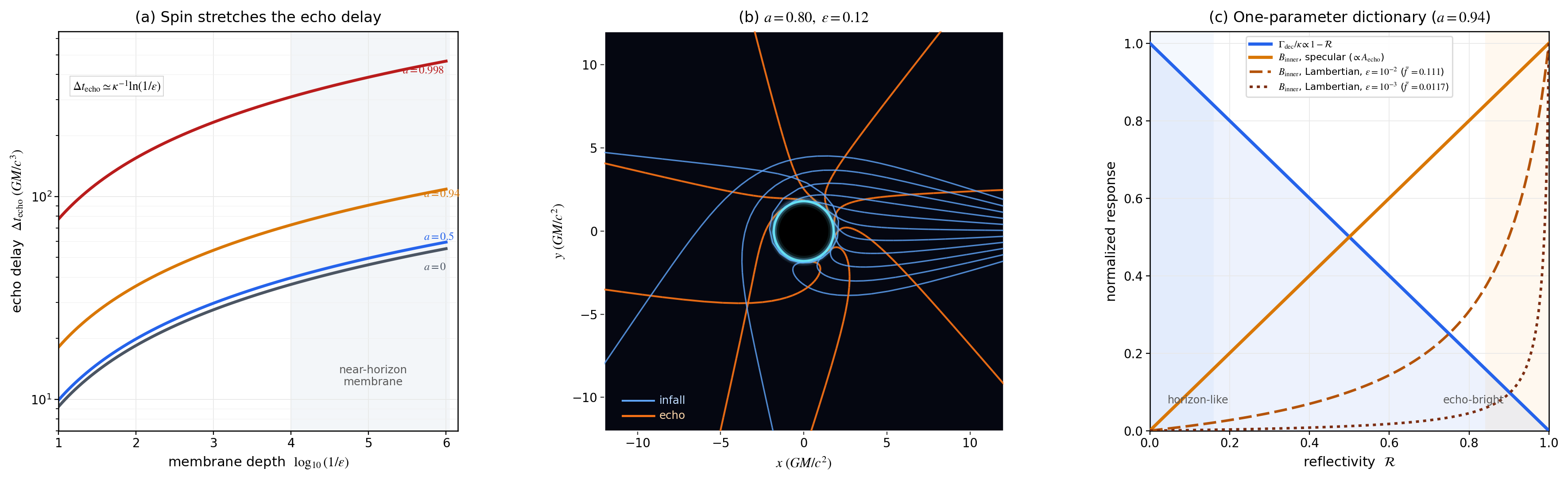}
\caption{The physical mapping from relativistic ray-tracing to the quantum-classical dictionary. (a) The logarithmic divergence of the macroscopic geometric delay near the theoretical \textcolor{black}{event} horizon \textcolor{black}{[scaling as $\kappa^{-1}\ln(1/\epsilon)$]}. (b) \textcolor{black}{An equatorial} ray-tracing diagram illustrating the geometric separatrix: the infalling trajectory (black) is destined to cross $r_+$, while the echo trajectory (orange) bounces off the membrane at $r_b$ and escapes \textcolor{black}{to infinity}. \textcolor{black}{(Every captublack ray within the photon shell is reflected exactly once before escaping).} (c) The robust one-parameter dictionary: The effective reflectivity $\mathcal{R}$ uniquely anti-correlates the microscopic quantum decoherence rate ($\Gamma_{\rm dec}/\kappa$) with macroscopic classical \textcolor{black}{imaging observables (such as the integrated inner shadow brightness $B_{\rm inner}$).} \textcolor{black}{Normalized $B_{\rm inner}$ profiles are compablack for both specular reflection and diffuse re-emission.}}
\label{fig:3}
\end{figure*}

Simultaneously, the reflected fraction $\mathcal{R}$ dynamically determines the macroscopic intensity of the classical echoes (e.g., the integrated inner shadow brightness $B_{\rm inner}$). This establishes a \textcolor{black}{monotonic, and hence invertible,} one-parameter correspondence \textcolor{black}{at fixed membrane position and reflection law [$B_{\rm inner}\propto\mathcal{R}$ for specular reflection and $B_{\rm inner}\propto\mathcal{R}(1-\eta_{\rm ret})/(1-\mathcal{R}\eta_{\rm ret})$ for diffuse re-emission; Eq.~(\ref{eq:10})]}: an object exhibiting strong quantum decoherence (horizon-like) must appear classically dark, whereas an object preserving external quantum coherence must manifest as a classically echo-bright geometry (Fig.~\ref{fig:3}c).

\par
\begin{figure*}[htbp]
\centering
\includegraphics[width=15cm]{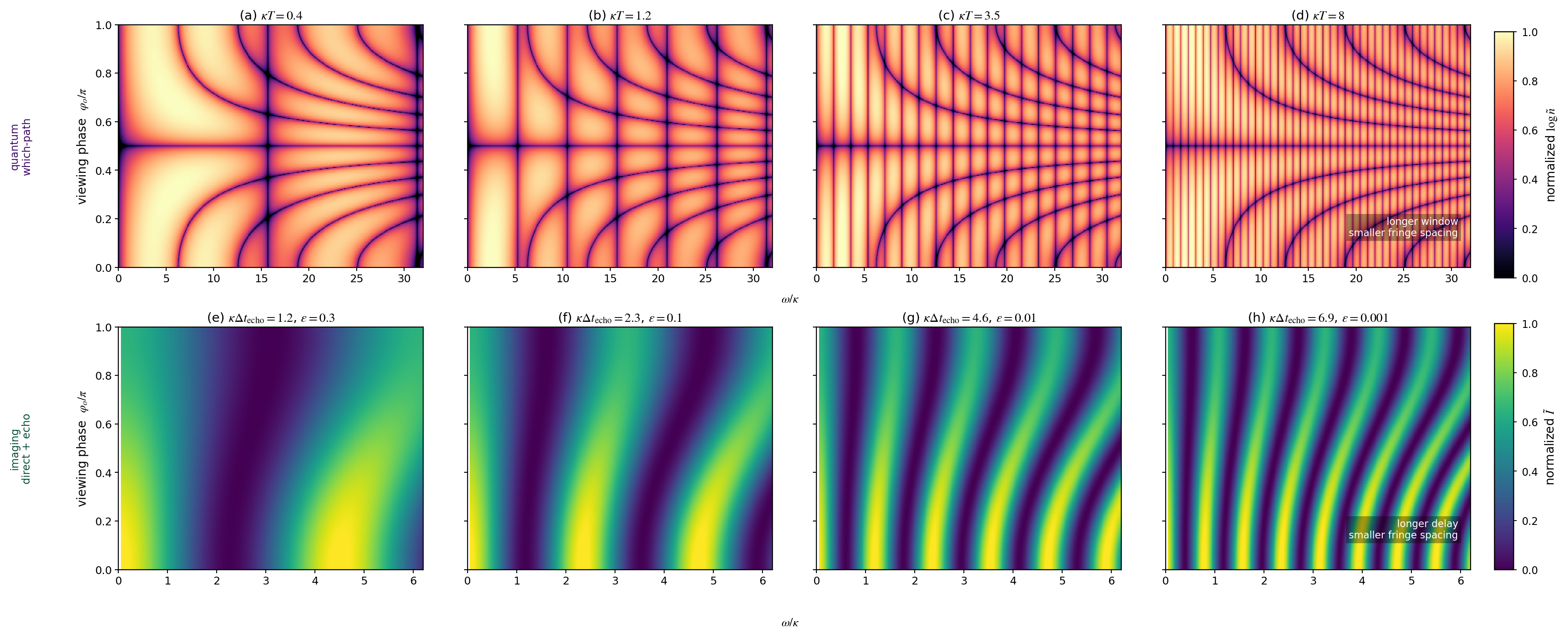}
\caption{The mirror spectra demonstrating the \textcolor{black}{structural analogy} between the quantum and classical domains. Top row (a-d): The quantum entangling-photon spectrum $\tilde{n}(\omega,\varphi_o)$ as derived in Eq.~(\ref{eq:8}), showing narrowing interference fringes for increasing quantum superposition times $\kappa T$. Bottom row (e-h): The classical \textcolor{black}{light-curve power} spectrum $\tilde{I}(\omega,\varphi_o)$ \textcolor{black}{of} Eq.~(\ref{eq:9}) \textcolor{black}{(echo amplitude $\mathcal{A}=0.78$, with a schematic dependence of the beaming and delay on the viewing phase)}, generated by varying the macroscopic echo delay $\Delta t_{\rm echo}$ (controlled by the boundary compactness $\epsilon$\textcolor{black}{; $\kappa\Delta t_{\rm echo}=\ln(1/\epsilon)$}). The frequency-domain fringe spacings \textcolor{black}{map onto each other ($T\leftrightarrow\Delta t_{\rm echo}$), linking} the classically observable \textcolor{black}{macroscopic} spectrum to the \textcolor{black}{microscopic} quantum state of the horizon.}
\label{fig:4}
\end{figure*}

Probing deeper into this microscopic-macroscopic mapping reveals a \textcolor{black}{structural analogy} in the frequency domain. In the quantum regime, the spectral-angular density $\tilde{n}(\omega)$ of the entangling soft photons, which dictates the fundamental loss of coherence over a time window $T$, is governed by the \textcolor{black}{thermal (Unruh)} response of the horizon\textcolor{black}{~\cite{Danielson:2022sga,Gralla:2023oya}; schematically,}
\begin{equation}
\tilde{n}(\omega) \propto \frac{1}{\omega} \sin^2\left(\frac{\omega d}{2}\right) \sin^2\left(\frac{\omega T}{2}\right) \coth\left(\frac{\pi\omega}{\kappa}\right).
\label{eq:8}
\end{equation}
Conversely, in the classical macroscopic regime, \textcolor{black}{the observed flux superimposes direct emission and its single delayed echo: $F(t)=L(t)+\mathcal{A}L(t-\Delta t_{\rm echo})$. Here, the echo amplitude $\mathcal{A}=\mathcal{P}\mathcal{R}$ includes the intercepted fraction $\mathcal{P}\le1$ and the reflectivity $\mathcal{R}$. Because specular rays return only once (Sec.~\ref{sec:3}), the power spectrum $\tilde{I}(\omega)=|\tilde F(\omega)|^2$ exhibits a periodic two-beam interference modulation}~\cite{Maggio:2017ivp}:
\begin{eqnarray}
&&\textcolor{black}{\frac{\tilde{I}(\omega)}{|\tilde L(\omega)|^2} = \left|1+\mathcal{A}\,e^{i\omega\Delta t_{\rm echo}}\right|^2}\nonumber\\
&&\textcolor{black}{= 1 + \mathcal{A}^2 + 2\mathcal{A}\cos(\omega \Delta t_{\rm echo})}.
\label{eq:9}
\end{eqnarray}
Comparing the oscillatory terms in Eq.~(\ref{eq:8}) and Eq.~(\ref{eq:9}) reveals a \textcolor{black}{formal} structural mapping. Despite their distinct physical origins, both domains exhibit \textcolor{black}{the same interference structure} (Fig.~\ref{fig:4}) \textcolor{black}{with respect to the conjugate Fourier frequency $\omega$ (the emitted mode frequency or the observed flux variability frequency, respectively).} The quantum superposition time $T$ acts as the \textcolor{black}{analog of} the macroscopic delay $\Delta t_{\rm echo}(\epsilon)$, strictly governing the \textcolor{black}{frequency-domain} fringe spacing. While not a physical equivalence, their structural correspondence is exact. $T$ and $\Delta t_{\rm echo}$ dictate the fringe spacings; the thermal factor $\coth(\pi\omega/\kappa)$ and the source spectrum $|\tilde L(\omega)|^2$ define the smooth envelopes. Crucially, the boundary couples to the spectra through complementary channels: the absorbed fraction ($1-\mathcal{R}$) dictates the quantum decoherence rate [Eq.~(\ref{eq:17})], while the reflected fraction ($\mathcal{R}$ via $\mathcal{A}$) determines the classical fringe depth. This analogy holds in the stationary, low-frequency regime ($\omega \lesssim \kappa$). Consequently, the power spectrum of the inner-shadow light curve serves as a macroscopic observational proxy for the boundary absorptivity $1-\mathcal{R}$, ultimately providing a handle to infer the unobservable horizon-induced quantum decoherence rate.

\section{Spatial Interferometric Signatures and VLBI Detectability}
\label{sec:3}
\par
Section~\ref{sec:2} established the fundamental dictionary mapping microscopic boundary conditions ($\mathcal{R}$, $\Gamma_{\rm dec}$) to delayed geometric echoes. To translate this theoretical framework into \textcolor{black}{macroscopically verifiable signatures}, we project the discrete time-domain echoes into the steady-state spatial and Fourier (visibility) domains. Here, we investigate how boundary reflections modify static interferometric signatures and define the ``Horizon Discovery Space'' to quantify the dynamic range requiblack for future VLBI arrays to constrain the quantum horizon \textcolor{black}{under realistic observational limits}. \textcolor{black}{Throughout, we bracket the unknown reflection physics with two limiting cases---specular reflection and diffuse (Lambertian) re-emission---and calibrate the effective signal-strength exponent $\Gamma(a,i)$ via full ray-tracing.}

\subsection{Inner Shadow Filling and Fourier Visibility Structure}
\label{sec:3-1}
\par
\begin{figure*}[htbp]
\centering
\includegraphics[width=15cm,height=3.5cm]{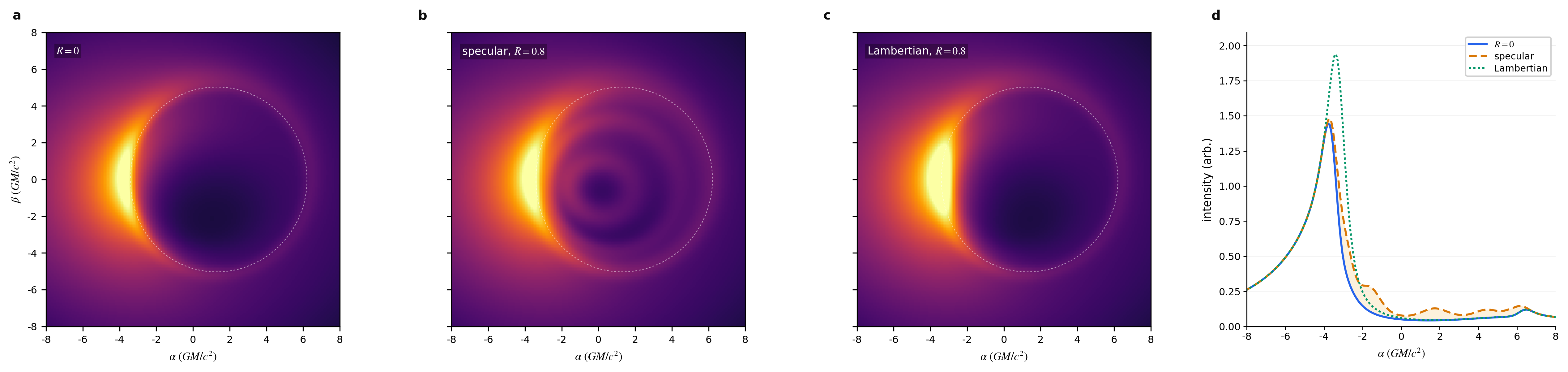}
\caption{Inner shadow filling as the spatial integration of temporal echoes ($a=0.94, i=40^\circ$\textcolor{black}{; membrane at $\epsilon=10^{-2}$, inside the photon shell}). (a) The standard Kerr black hole ($\mathcal{R}=0$) naturally \textcolor{black}{exhibits} a deep central depression due to irreversible photon capture. \textcolor{black}{(b) Specular membrane with $\mathcal{R}=0.8$: every captublack ray is reflected exactly once and escapes, filling the inner shadow with nested reflection rings; the excess is exactly linear in $\mathcal{R}$ [$\eta_{\rm ret}=0$ in Eq.~(\ref{eq:10})]. (c) Lambertian membrane with $\mathcal{R}=0.8$: only the escape-cone fraction $\bar f=0.11$ of the re-emitted light escapes per re-processing [$\eta_{\rm ret}=1-\bar f$], and the escaping light is Doppler-beamed into a bright crescent along the prograde edge of the critical curve (dashed). (d) Horizontal cut at $\beta\simeq0$; within $3\,GM/c^2$ of the image centre the flux increases by factors of $1.94$ (specular) and $1.48$ (Lambertian). All panels share the same finite-resolution beam and colour scale.}}
\label{fig:5}
\end{figure*}

The primary physical mechanism bridging the delayed temporal echoes to steady-state spatial imaging is the continuous temporal integration of reflected null trajectories. Assuming a stationary, optically thin accretion flow, the discrete temporal echoes map ergodically into a static spatial intensity distribution. For a line of sight intersecting the lensed inner shadow, a true Kerr event horizon ($\mathcal{R}=0$) completely absorbs the infalling geodesics, rendering the central depression strictly dark (Fig.~\ref{fig:5}a). \textcolor{black}{For a reflective membrane ($\mathcal{R}>0$), the returned intensity depends on the microscopic reflection law, which we bracket by two limiting cases. (i) \textit{Specular reflection} ($p_r\to-p_r$ at $r_b$, conserving $E$, $L$, and $Q$): a ray inside the critical curve lacks a radial turning point outside the horizon [$R(r)>0$ for all $r>r_+$] and thus escapes to infinity after exactly one reflection; repeated bounces between $r_b$ and the potential barrier do not occur for null geodesics. Conversely, rays outside the critical curve never reach the membrane, provided it lies inside the photon shell ($\epsilon<\epsilon_{\rm ph}(a)\equiv r_{\rm ph}/r_+-1$, where $r_{\rm ph}$ is the prograde circular photon orbit radius). (ii) \textit{Lambertian (diffuse) re-emission} in the rest frame corotating with the horizon angular velocity $\Omega_H$: only a fraction $\bar f$ of the re-emitted photons falls within the local escape cone to reach infinity. The remaining fraction $\eta_{\rm ret}=1-\bar f$ (which includes all negative-energy directions if the membrane lies inside the ergoregion) falls back onto the membrane and is repeatedly re-processed~\cite{Broderick:2005xa,Lu:2017vdx}.}

\par
\textcolor{black}{Summing these successive re-processings yields a geometric series. By invoking Liouville's theorem ($I_\nu/\nu^3$ is conserved along rays)~\cite{Lindquist:1966igj} and treating the diffuse re-processing in a one-zone approximation,} the accumulated pixel intensity $I_{\rm pix}$ at the observer's screen \textcolor{black}{is given by} \textcolor{black}{the unified form}\textcolor{black}{:}
\begin{eqnarray}
I_{\rm pix}(\mathcal{R}) &=& I_{\rm direct} + I_{\rm inc}\,\mathcal{R}(1-\eta_{\rm ret})\sum_{k=0}^{\infty}\left(\mathcal{R}\eta_{\rm ret}\right)^k \nonumber\\
&=& I_{\rm direct} + I_{\rm inc}\,\frac{\mathcal{R}(1-\eta_{\rm ret})}{1-\mathcal{R}\eta_{\rm ret}},
\label{eq:10}
\end{eqnarray}
\textcolor{black}{where $I_{\rm direct}$ is the unreflected foreground emission and $I_{\rm inc}$ is the blackshift-weighted incident emission (for diffuse re-emission, this is the Doppler-weighted average over all incident directions). The specular limit corresponds to $\eta_{\rm ret}=0$, yielding an exactly linear filling in $\mathcal{R}$~\cite{Carballo-Rubio:2022aed}. The Lambertian limit corresponds to $\eta_{\rm ret}=1-\bar f(\epsilon;a)$, with $\bar f$ derived in Sec.~\ref{sec:3-2}. The numerator $\mathcal{R}(1-\eta_{\rm ret})$ ensures the series sums to unity for $\mathcal{R}=1$, thereby conserving the radiation energy measublack in the membrane rest frame (for $a\neq0$, this corresponds to $E-\Omega_H L$ rather than the Killing energy $E$).} Consequently, the excess continuous brightness filling the inner shadow, \textcolor{black}{$B_{\rm inner}\propto\mathcal{R}(1-\eta_{\rm ret})/(1-\mathcal{R}\eta_{\rm ret})$}, constitutes a robust, macroscopically observable signature of the horizon boundary condition\textcolor{black}{; it is a monotonic function of $\mathcal{R}$ and coincides with the fractional filling $b_{\rm inner}$ used in Sec.~\ref{sec:4}}. \textcolor{black}{Figure~\ref{fig:5} illustrates both limits for $\mathcal{R}=0.8$ with a membrane at $\epsilon=10^{-2}$, located inside the photon shell ($r_b=1.355<r_{\rm ph}=1.425$ for $a=0.94$). The specular membrane fills the inner shadow with a nested set of reflection rings converging onto the critical curve (Fig.~\ref{fig:5}b)~\cite{deKleuver:2023nwv}, achieving a mean core brightness contrast of $C=0.21$ relative to the photon ring (the contrast $C$ is defined in Sec.~\ref{sec:3-2}). The Lambertian membrane leaves the core darker ($C=0.036$), as only $\bar f=0.11$ of the light escapes per bounce. Escaping photons are Doppler-beamed into a bright crescent along the prograde edge of the critical curve (Fig.~\ref{fig:5}c), peaking where the corotating membrane's kinematic factor $[\gamma_v(1-\Omega_H\xi)]^{-4}$ is maximized ($\gamma_v$ is the Lorentz factor of the membrane relative to the zero-angular-momentum frame). Within $3\,GM/c^2$ of the image centre, the flux increases by factors of $1.94$ (specular) and $1.48$ (Lambertian) relative to $\mathcal{R}=0$ (Fig.~\ref{fig:5}d).}

\par
\begin{figure*}[htbp]
\centering
\includegraphics[width=15cm,height=4.5cm]{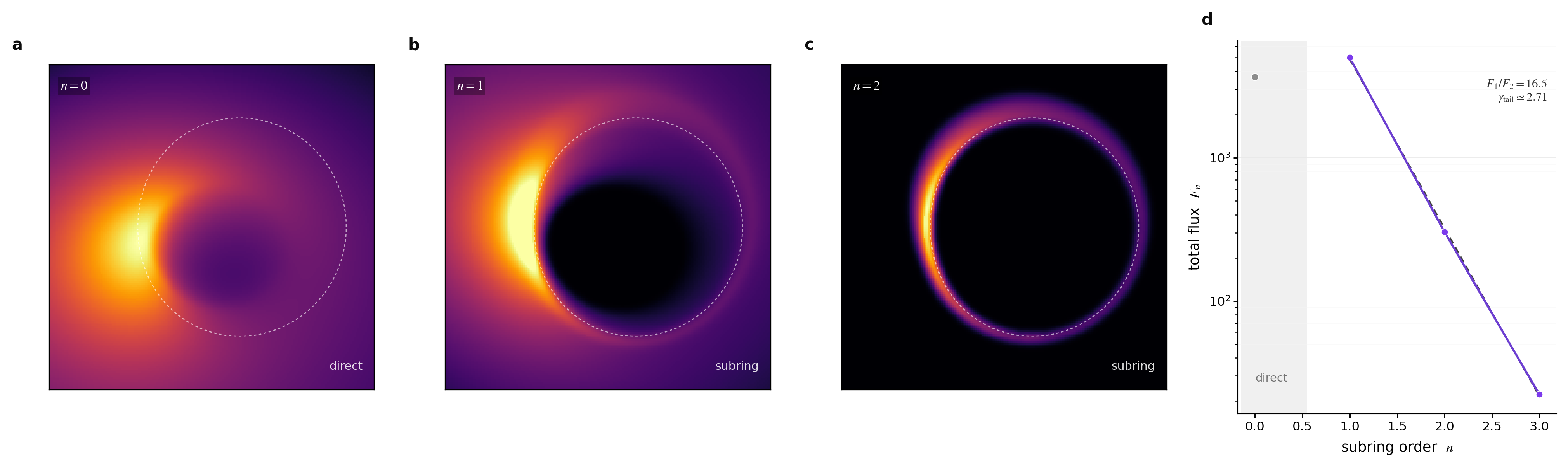}
\caption{Photon subring decomposition illustrating the self-similar lensing nature independent of the boundary parameter $\mathcal{R}$ \textcolor{black}{($\mathcal{R}=0$ image; $a=0.94$, $i=40^\circ$, $\epsilon=10^{-2}$)}. Panels (a-c) isolate the direct emission ($n=0$) and the subsequent strongly lensed photon subrings ($n=1, 2$)\textcolor{black}{, classified by the number $n$ of equatorial-plane crossings (for the geometrically thick flow the $n=1$ map also contains the direct emission from the part of the flow below the equatorial plane); the dashed curve is the analytic critical curve}. Panel (d) demonstrates that the integrated total flux $F_n$ \textcolor{black}{follows} the Lyapunov demagnification law $F_n \propto e^{-\gamma n}$ dictated exclusively by the background Kerr metric \textcolor{black}{($F_1/F_2=16.5$, $\gamma_{\rm tail}\simeq2.71$; the last point collects $n\ge3$)}, establishing a boundary-independent calibration reference\textcolor{black}{: since the membrane lies inside the photon shell, every pixel outside the critical curve is identical for all $\mathcal{R}$}.}
\label{fig:6}
\end{figure*}

To observationally isolate this boundary-dependent inner shadow filling from intrinsic metric lensing uncertainties, we utilize the mathematically pristine photon subrings as a \textcolor{black}{robust} spacetime baseline. Formed by rays executing $n \ge 1$ half-orbits around the unstable spherical photon shell entirely outside the \textcolor{black}{membrane}, these subrings are immune to $\mathcal{R}$ \textcolor{black}{as long as the membrane lies inside the photon shell, $\epsilon<\epsilon_{\rm ph}(a)$ ($\epsilon_{\rm ph}=0.063$ for $a=0.94$): reflected light re-emerges exclusively inside the critical curve, leaving every exterior pixel mathematically identical across all $\mathcal{R}$, as verified pixel-by-pixel in both limits}. The geometric demagnification of successive subrings is governed exclusively by the Lyapunov exponent $\gamma$ of the radial effective potential, which characterizes the chaotic divergence of near-critical geodesics. This imposes a universal, metric-dependent exponential decay on the integrated subring flux $F_n$:
\begin{equation}
F_n \propto e^{-\gamma n}.
\label{eq:11}
\end{equation}
Figure~\ref{fig:6} provides the numerical subring decomposition \textcolor{black}{of the $\mathcal{R}=0$ image by the number $n$ of equatorial-plane crossings} ($n=0, 1, 2$), visually demonstrating that the higher-order subrings rapidly converge to the analytic critical curve. Figure~\ref{fig:6}d confirms the \textcolor{black}{exponential} Lyapunov scaling, yielding \textcolor{black}{$\gamma_{\rm tail} \approx 2.71$ from $n=1$--$3$} for this specific Kerr geometry, independent of the boundary physics introduced at $r_b$. \textcolor{black}{Note that for geometrically thick flows, the $n=1$ map additionally contains direct emission from below the equatorial plane, which is why $F_1>F_0$.}

\par
\begin{figure*}[htbp]
\centering
\includegraphics[width=15cm]{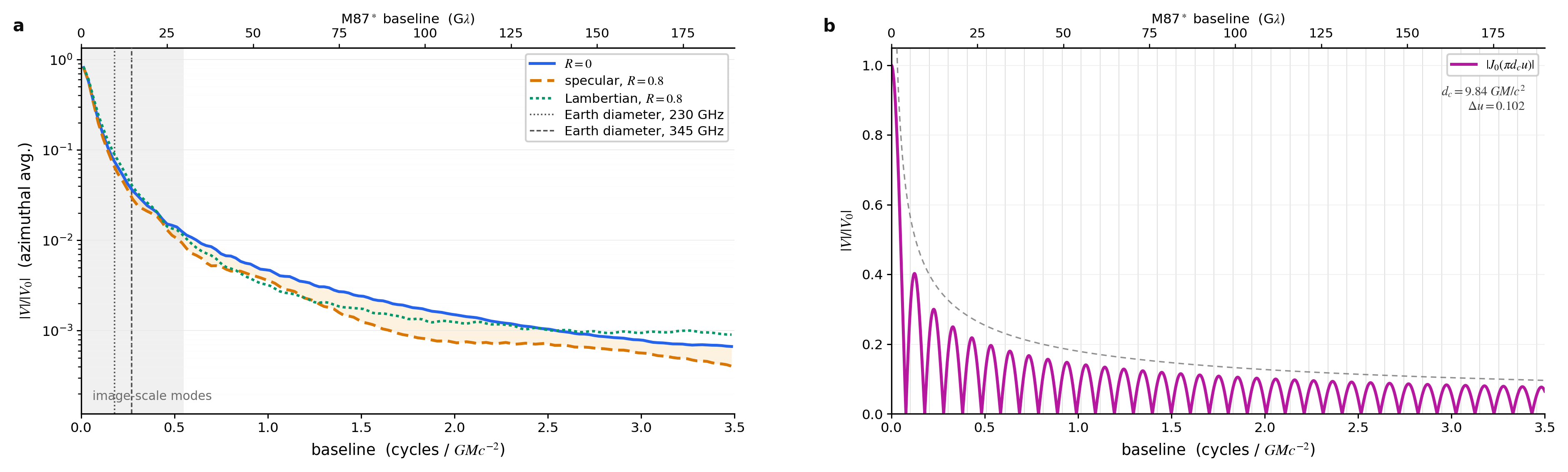}
\caption{Interferometric visibility demonstrating the fundamental decoupling of boundary physics and spacetime geometry in the Fourier domain. \textcolor{black}{(a) Azimuthally averaged visibility amplitude of the $\mathcal{R}=0$ image and of the specular and Lambertian images with $\mathcal{R}=0.8$ (Fig.~\ref{fig:5}; $a=0.94$, $i=40^\circ$, $\epsilon=10^{-2}$). At low spatial frequencies (short baselines), the inner-shadow filling modifies the macroscopic image structure and alters the visibility amplitude envelope; vertical lines mark the longest Earth baselines for M87$^\ast$ at 230 and 345~GHz, and the top axis gives the baseline in G$\lambda$ for M87$^\ast$.} (b) At high spatial frequencies (long baselines), the metric-governed sharp photon ring dominates the signal, maintaining the universal $|J_0|$ ringing \textcolor{black}{[analytic $|J_0(\pi d_c u)|$ of the critical curve]} with a strict period $\Delta u \simeq 0.102 \, (\text{cycles} / GM c^{-2})$ \textcolor{black}{that is} completely independent of $\mathcal{R}$.}
\label{fig:7}
\end{figure*}

This theoretical dichotomy (an $\mathcal{R}$-dependent inner shadow enveloped by an \textcolor{black}{$\mathcal{R}$-independent photon-ring geometry}) manifests in the complex visibility domain $\mathcal{V}(\mathbf{u})$ measublack by VLBI arrays. According to the van Cittert-Zernike theorem, the spatial Fourier transform of the infinitesimally thin photon subrings, characterized by an asymptotic celestial diameter $d_c$, imprints a universal damped oscillation on the visibility amplitude plane~\cite{Johnson:2019ljv}\textcolor{black}{:}
\begin{equation}
|\mathcal{V}(u)| \propto \big| J_0(\pi d_c u) \big|,
\label{eq:12}
\end{equation}
where $u = |\mathbf{u}|$ is the dimensionless baseline length (spatial frequency) and $J_0$ is the zeroth-order Bessel function of the first kind. Figure~\ref{fig:7} \textcolor{black}{clearly} demonstrates this spatial-frequency decoupling. \textcolor{black}{At short baselines ($u\lesssim0.5$ cycles$/GMc^{-2}$, corresponding to $\lesssim27\,{\rm G}\lambda$ for M87$^\ast$; Fig.~\ref{fig:7}a), the inner-shadow filling induced by $\mathcal{R}=0.8$ alters the normalized visibility amplitude by a mean factor of $0.87$ (specular) or $1.09$ (Lambertian) relative to $\mathcal{R}=0$, yielding a direct measurement of the macroscopic echo intensity. The ray-traced amplitude also exhibits changes at long baselines (mean factors of $0.60$ and $0.92$ for $1.5\le u\le3$) as reflected rings converge onto the critical curve. However, the underlying photon-ring geometry remains strictly invariant: the diameter $d_c$, and thus the universal periodicity $\Delta u\simeq1/d_c$ of the $|J_0|$ ringing (Fig.~\ref{fig:7}b), is fixed entirely by the metric.} Thus, a critical observational strategy emerges: \textcolor{black}{the ring diameter and periodicity measublack on} ultra-long baselines calibrate the background Kerr metric ($d_c, \gamma$), thereby establishing a rigid mathematical reference to robustly extract the boundary parameter $\mathcal{R}$ from the \textcolor{black}{visibility amplitudes, most directly through their short-baseline deviations}.

\par
\textcolor{black}{To observationally evaluate the feasibility of this Fourier-domain separation, we must quantify the requiblack instrumental limits and astrophysical contaminations. For M87$^\ast$ ($GM/c^2D\simeq3.8\,\mu$as), $u=1$ cycle$/GMc^{-2}$ equates to $54\,{\rm G}\lambda$. Earth-diameter baselines therefore reach $u\lesssim0.18$ at 230~GHz and $u\lesssim0.27$ at 345~GHz [vertical lines in Fig.~\ref{fig:7}(a)]. Across the $0.05\le u\le0.27$ range, the inner-shadow filling ($\mathcal{R}=0.8$, $\epsilon=10^{-2}$) alters the normalized visibility amplitude by $\Delta|V|/|V_0|\simeq0.006$--$0.017$ (specular) and $0.007$--$0.036$ (Lambertian). This translates to a deviation of $3.5$--$15\%$ (median $11\%$) and $11$--$21\%$ (median $17\%$) relative to the baseline $\mathcal{R}=0$ amplitude. Achieving a $3\sigma$ detection on a single visibility thus requires an interferometric signal-to-noise ratio (SNR) of $\simeq28$ (specular) or $\simeq17$ (Lambertian) at the median change. For the specular case, this requirement scales approximately as $1/\mathcal{R}$ and improves as $N_{\rm vis}^{-1/2}$ when combining $N_{\rm vis}$ independent visibilities. Current and next-generation Event Horizon Telescope (ngEHT) configurations involving sensitive anchor stations (e.g., ALMA) readily achieve SNR values of $\mathcal{O}(10^2)$, comfortably exceeding this thermal-noise threshold. The primary limitation is station-gain calibration; its $\sim10\%$ amplitude uncertainty is comparable to the signal and must be mitigated via closure amplitudes or self-calibration. Crucially, unlike photon ring signatures [Fig.~\ref{fig:7}(b)] that require space-VLBI baselines, this inner-shadow signal is carried by image-scale modes natively matched to the dense $(u,v)$-coverage of Earth-based arrays.}

\par
\textcolor{black}{Furthermore, this low-baseline signature is intrinsically resilient to standard astrophysical complexities:}
\begin{itemize}
    \item \textcolor{black}{\textbf{Non-stationarity:} Although accretion flows are turbulent, the macroscopic inner shadow is a stable geometric structure. Time-averaged visibilities robustly recover the mean reflected component, provided the flow is statistically stationary over the echo delay (days to weeks for M87$^\ast$; Sec.~\ref{sec:3-2}), making multi-epoch averaging highly beneficial.}
    \item \textcolor{black}{\textbf{Interstellar Scattering:} The scattering screen (e.g., for Sgr A$^\ast$) acts mathematically as a Gaussian low-pass filter in the visibility domain. While it severely suppresses high-frequency photon ring fringes [Fig.~\ref{fig:7}(b)], it leaves the crucial low-baseline inner-shadow signature [Fig.~\ref{fig:7}(a)] largely unattenuated; refractive substructure primarily adds noise on long baselines. Thus, probing the inner shadow provides a highly scattering-immune observational channel.}
    \item \textcolor{black}{\textbf{Refraction:} At observation frequencies of 230~GHz and 345~GHz, the near-horizon accretion plasma is optically thin with a refractive index $n \approx 1$. Dispersive refraction is negligible and cannot systematically mimic the geometry-governed visibility deviations induced by a reflective membrane.}
\end{itemize}
\par
\textcolor{black}{Therefore, the boundary signal (low-frequency visibility amplitude) can be structurally decoupled from the spacetime geometry (ring diameter and ringing period) using next-generation arrays, leaving amplitude calibration as the dominant systematic uncertainty.}

\subsection{Dynamic Range Constraints in the Parameter Space}
\label{sec:3-2}
\par
To rigorously evaluate the real-world detectability of these quantum boundary features \textcolor{black}{under realistic instrumental limits}, we define a unified parameter space spanned by the microscopic reflectivity $\mathcal{R}$ and the dimensionless membrane compactness $\epsilon$. The macroscopic observability of the boundary is fundamentally constrained by two coupled quantities: the effective observable signal strength $C(\mathcal{R}, \epsilon)$ (characterizing the relative flux contrast of the echoes\textcolor{black}{, defined as the mean brightness excess in the inner-shadow core---pixels inside the critical curve and more than $1\,GM/c^2$ away from it---in units of the mean $\mathcal{R}=0$ brightness of the photon-ring band $|\Delta b|<0.5\,GM/c^2$, which dictates the dynamic range needed for detection}) and the geometric echo delay time $\Delta t_{\rm echo}$\textcolor{black}{. Following Eq.~(\ref{eq:10}), our ray-tracing calibration (detailed below) is parameterized by:}
\begin{eqnarray}
{\color{black}C(\mathcal{R}, \epsilon; a, i)} &{\color{black}=}& {\color{black}w(a,i)\,\frac{\mathcal{R}(1-\eta_{\rm ret})}{1-\mathcal{R}\eta_{\rm ret}}, \qquad \Gamma\equiv\frac{\partial\ln C}{\partial\ln\epsilon},} \label{eq:13} \\
\Delta t_{\rm echo}(\epsilon, a) &\simeq& \frac{1}{\kappa} \ln\left(\frac{1}{\epsilon}\right) t_g, \label{eq:14}
\end{eqnarray}
where $t_g = GM/c^3$ is the gravitational timescale\textcolor{black}{, $\eta_{\rm ret}=0$ and $w=w_s(a,i)$ for a specular membrane, and $\eta_{\rm ret}=1-\bar f(\epsilon;a)$ and $w=N(a,i)$ for a Lambertian membrane. Locally, $C\propto\mathcal{R}\,\epsilon^{\Gamma}$, but the effective exponent $\Gamma$ is not a universal constant: it is determined by the reflection law through Liouville's theorem, the escape cone, and the blackshift factor, as detailed below.}

\par
\textcolor{black}{Since $I_\nu/\nu^3$ is conserved along rays~\cite{Lindquist:1966igj} (and $I/\nu^4$ for bolometric intensity), the returned intensity is dictated by the constants of motion $(E,L,Q)$ alongside the frequency shifts at emission and reflection. \textit{Specular limit.} Reflection preserves $E, L, Q$ and is perfectly elastic in any stationary frame. An infalling photon arriving from infinity with a local blackshift $\nu_{\rm loc}/\nu_\infty=\gamma_v(1-\Omega_H\xi)/\zeta$ (where $\zeta=\sqrt{\Sigma\Delta/A}$ is the lapse) departs with the exact inverse blackshift. Furthermore, the mapping $n_r\to-n_r$ bijectively connects the capture cone arriving from infinity onto the escape cone~\cite{Synge:1966okc}. Both cones shrink proportionally to $\zeta^2$ as $\epsilon\to0$, meaning they shrink in tandem and no flux is lost: the reflected intensity strictly equals $\mathcal{R}$ times the incident emission, independent of $\epsilon$. Consequently, $\Gamma=0$ up to $\mathcal{O}(\epsilon)$ corrections; our ray-traced images confirm a local exponent $|\Gamma|<4\times10^{-3}$ for $\epsilon\le10^{-2}$ (and $|\Gamma|<0.013$ up to $\epsilon=3\times10^{-2}$) across all $(a,i)$. \textit{Lambertian limit.} Re-emitted directions are locally randomized, and the escaping fraction equals the \'etendue of the escape cone,}
\begin{equation}
f(\theta) = \frac{1}{\pi}\int_{\rm esc} dn_\theta\, dn_\phi = \frac{\zeta^2}{\pi\,\varpi\sqrt{\Sigma}} \int_{{\rm Sh}(\theta)} \frac{d\xi\, d\hat p_\theta}{\gamma_v^3\,(1-\Omega_H\xi)^3},
\label{eq:13b}
\end{equation}
\textcolor{black}{evaluated at $r_b$, where the integral spans the Bardeen shadow region parameterized by $(\xi,\hat p_\theta\equiv p_\theta/E)$ observed from latitude $\theta$, and $\varpi=\sqrt{A/\Sigma}\sin\theta$. In steady state, the kinematic incident blackshift ($\zeta^{-4}$) and outgoing blackshift ($\zeta^{4}$) strictly cancel. The rotating membrane's Doppler factor $[\gamma_v(1-\Omega_H\xi)]^{\mp4}$ does not cancel, but it is independent of $\epsilon$ to $\mathcal{O}(\epsilon)$ and is entirely absorbed into the prefactor $N(a,i)$. For $a=0$, this kinematic balance recovers Kirchhoff's law; for $a\neq0$ and $\mathcal{R}\to1$, the membrane can return more Killing energy than it receives, effectively extracting rotational energy akin to the ergoregion instability of spinning horizonless objects~\cite{Maggio:2017ivp}. The macroscopic suppression originates exclusively from the collapsing escape cone \'etendue ($\zeta^2\propto\Delta(r_b)\propto\epsilon$), yielding an asymptotic exponent of exactly unity. For $a=0$, Eq.~(\ref{eq:13b}) matches the exact analytic result $f=\sin^2\psi_c=(27/4)\,\epsilon/(1+\epsilon)^3$~\cite{Synge:1966okc,Lu:2017vdx}; for $a\neq0$, it aligns with our Monte Carlo sampling of re-emitted directions within statistical errors ($\le1.3\sigma$ across 24 test cases), perfectly consistent with the vanishing escape probability near sub-extremal horizons~\cite{Ogasawara:2019mir}.}

\par
\begin{figure*}[htbp]
\centering
\includegraphics[width=15cm]{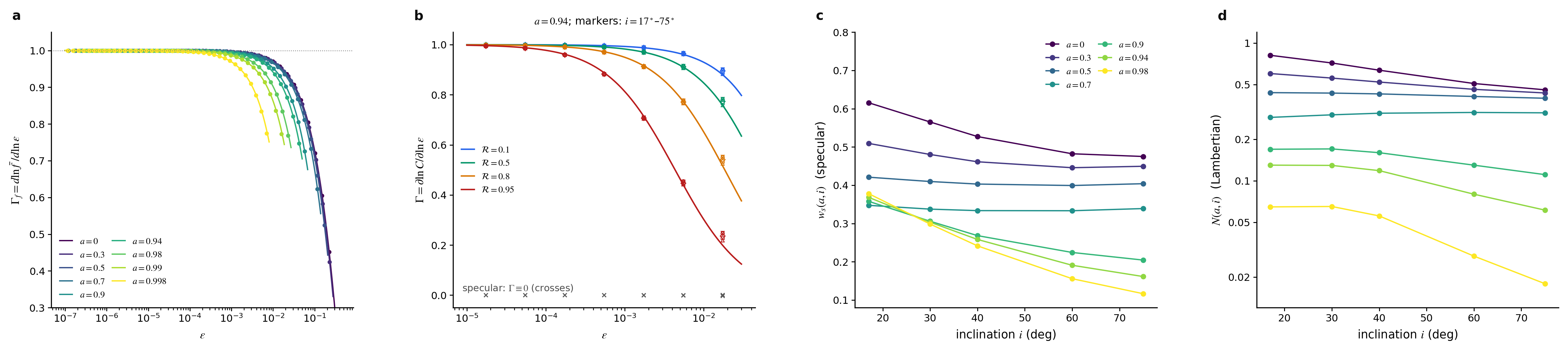}
\caption{\textcolor{black}{Calibration of the exponent $\Gamma(a,i)$. (a) Local exponent $\Gamma_f=d\ln\bar f/d\ln\epsilon$ of the Lambertian escape fraction for $0\le a\le0.998$ (points: direct quadrature of Eq.~(\ref{eq:13b}); curves: fit of Eq.~(\ref{eq:13c}) with Table~\ref{tab:gamma_fit}); $\Gamma_f\to1$ as $\epsilon\to0$. (b) Image-level $\Gamma=\partial\ln C/\partial\ln\epsilon$ for $a=0.94$, four reflectivities and five inclinations $17^\circ\le i\le75^\circ$ (open markers), compablack with the inclination-independent formula (curves); crosses: specular membrane, $\Gamma=0$. (c) Specular prefactor $w_s(a,i)$. (d) Lambertian prefactor $N(a,i)$.}}
\label{fig:gamma}
\end{figure*}

\textcolor{black}{For the Lambertian limit, we fit the area-averaged escape fraction as}
{\color{black}
\begin{eqnarray}
&&\bar f(\epsilon;a) = K(a)\,\epsilon\left(1+\frac{q\,\epsilon}{\epsilon_c}\right)^{-1/q}, \nonumber\\
&&\Gamma = \Gamma_f\,\frac{1-\mathcal{R}}{1-\mathcal{R}+\mathcal{R}\bar f}, \qquad \Gamma_f = 1-\frac{\epsilon/\epsilon_c}{1+q\,\epsilon/\epsilon_c},
\label{eq:13c}
\end{eqnarray}
}
\textcolor{black}{with parameters $(K,\epsilon_c,q)$ specified in Table~\ref{tab:gamma_fit}. This fit is exact for $a=0$ ($K=27/4$, $\epsilon_c=q=1/3$) and accurate to $3\times10^{-3}$ for $0\le a\le0.998$ across $10^{-7}\le\epsilon\le0.8\,\epsilon_{\rm ph}(a)$; the coefficient is approximated by $K(a)=\tfrac{27}{4}x^{-0.9787}[1-0.1903(1-x)-0.0135(1-x)^2]$ with $x=(r_+-r_-)/r_+=4\kappa$ (relative error $10^{-4}$). Equation~(\ref{eq:13c}) is intrinsically inclination-independent, as inclination only enters through the prefactor $N(a,i)$. Table~\ref{tab:gamma_ai} and Fig.~\ref{fig:gamma} compare it directly with image-level exponents from full ray-tracing spanning $0\le a\le0.98$, $17^\circ\le i\le75^\circ$, and $10^{-5}\le\epsilon\le3\times10^{-2}$: the analytical and numerical results agree within $0.008$ for $\epsilon\le0.1\,\epsilon_{\rm ph}$ and $0.045$ for $\epsilon\le0.5\,\epsilon_{\rm ph}$. Closer to the photon shell (e.g., $a=0.98$, $\epsilon=3\times10^{-2}\simeq0.9\,\epsilon_{\rm ph}$), the deviation slightly grows to $0.13$ because $N$ itself begins to acquire $\epsilon$-dependence. The exponent $\Gamma_f\to1$ for $\epsilon\ll\epsilon_c(a)$, where $\epsilon_c$ decreases from $1/3$ ($a=0$) to $0.032$ ($a=0.998$). Meanwhile, the multiple-return factor $(1-\mathcal{R})/(1-\mathcal{R}+\mathcal{R}\bar f)$ marginally lowers $\Gamma$ at large $\mathcal{R}$. Finally, altering the membrane rest frame (corotating vs. zero-angular-momentum) changes $\bar f$ by $13\%$ at $\epsilon=10^{-2}$ and by $1.4\%$ at $\epsilon=10^{-3}$ for $a=0.94$; both frames converge identically as $\epsilon\to0$.}
\begin{table*}[t]
\color{black}
\caption{Lambertian membrane corotating with the horizon ($\Omega_H$): parameters of the escape-fraction fit $\bar f(\epsilon;a)=K\epsilon\,(1+q\epsilon/\epsilon_c)^{-1/q}$ ($q\to0$: $K\epsilon e^{-\epsilon/\epsilon_c}$) and the geometric exponent $\Gamma_f=d\ln\bar f/d\ln\epsilon=1-(\epsilon/\epsilon_c)/(1+q\epsilon/\epsilon_c)$. Fit range $10^{-7}\le\epsilon\le0.8\,\epsilon_{\rm ph}$; maximum relative error 2.6e-03. Closed form: $K(a)=\frac{27}{4}x^{-0.9787}[1-0.1903(1-x)-0.0135(1-x)^2]$, $x=(r_+-r_-)/r_+=4\kappa$ (max. error 1e-04).}
\label{tab:gamma_fit}
\begin{ruledtabular}
\begin{tabular}{ccccccccc}
$a$ & $\kappa$ & $\epsilon_{\rm ph}$ & $K$ & $\epsilon_c$ & $q$ & $\Gamma_f(10^{-2})$ & $\Gamma_f(10^{-3})$ & $\Gamma_f(10^{-4})$ \\
\hline
0 & 0.2500 & 0.5000 & 6.750 & 0.3333 & 0.333 & 0.970 & 0.997 & 1.000 \\
0.3 & 0.2441 & 0.3460 & 6.878 & 0.3289 & 0.305 & 0.970 & 0.997 & 1.000 \\
0.5 & 0.2321 & 0.2579 & 7.160 & 0.3179 & 0.254 & 0.969 & 0.997 & 1.000 \\
0.7 & 0.2083 & 0.1745 & 7.810 & 0.2911 & 0.167 & 0.966 & 0.997 & 1.000 \\
0.9 & 0.1518 & 0.0849 & 10.155 & 0.2096 & 0.000 & 0.952 & 0.995 & 1.000 \\
0.94 & 0.1272 & 0.0627 & 11.814 & 0.1700 & 0.000 & 0.941 & 0.994 & 0.999 \\
0.98 & 0.0830 & 0.0338 & 17.220 & 0.1024 & 0.000 & 0.902 & 0.990 & 0.999 \\
0.99 & 0.0618 & 0.0233 & 22.500 & 0.0728 & 0.000 & 0.863 & 0.986 & 0.999 \\
0.998 & 0.0297 & 0.0101 & 44.589 & 0.0323 & 0.000 & --- & 0.969 & 0.997 \\
\end{tabular}
\end{ruledtabular}
\end{table*}
\begin{table*}[t]
\color{black}
\caption{Image-level exponent $\Gamma(a,i)\equiv\partial\ln C/\partial\ln\epsilon$ for a Lambertian membrane with $\mathcal{R}=0.8$, at $\epsilon\simeq3\times10^{-3}$ / $3\times10^{-4}$ (log-centblack chords $10^{-3}$--$10^{-2}$ and $10^{-4}$--$10^{-3}$), compablack with the inclination-independent formula $\Gamma=\Gamma_f(1-\mathcal{R})/(1-\mathcal{R}+\mathcal{R}\bar f)$ with Table~\ref{tab:gamma_fit} (last column); maximum difference 0.022. For a specular membrane the local exponent satisfies $|\Gamma|<0.004$ for $\epsilon\le10^{-2}$ and $|\Gamma|<0.013$ up to $3\times10^{-2}$, for every $(a,i)$.}
\label{tab:gamma_ai}
\begin{ruledtabular}
\begin{tabular}{ccccccc}
$a$ & $i=17^\circ$ & $i=30^\circ$ & $i=40^\circ$ & $i=60^\circ$ & $i=75^\circ$ & formula \\
\hline
0 & 0.899 / 0.988 & 0.899 / 0.988 & 0.899 / 0.988 & 0.899 / 0.989 & 0.899 / 0.989 & 0.899 / 0.988 \\
0.3 & 0.897 / 0.988 & 0.897 / 0.988 & 0.897 / 0.988 & 0.897 / 0.988 & 0.897 / 0.988 & 0.897 / 0.988 \\
0.5 & 0.894 / 0.988 & 0.893 / 0.988 & 0.893 / 0.988 & 0.893 / 0.988 & 0.893 / 0.988 & 0.894 / 0.988 \\
0.7 & 0.886 / 0.987 & 0.885 / 0.987 & 0.885 / 0.987 & 0.885 / 0.987 & 0.885 / 0.987 & 0.885 / 0.987 \\
0.9 & 0.859 / 0.983 & 0.856 / 0.983 & 0.855 / 0.983 & 0.856 / 0.983 & 0.859 / 0.983 & 0.856 / 0.983 \\
0.94 & 0.843 / 0.980 & 0.838 / 0.980 & 0.835 / 0.980 & 0.839 / 0.980 & 0.843 / 0.980 & 0.837 / 0.980 \\
0.98 & 0.801 / 0.971 & 0.790 / 0.971 & 0.783 / 0.971 & 0.794 / 0.971 & 0.800 / 0.971 & 0.780 / 0.971 \\
\end{tabular}
\end{ruledtabular}
\end{table*}

\par
By analytically eliminating the unobservable microscopic parameter $\epsilon$ using Eq.~(\ref{eq:14})\textcolor{black}{, $\epsilon=\exp(-\kappa\Delta t_{\rm echo}/t_g)$, the two limits reveal distinct macroscopic consequences: the specular signal is entirely independent of the delay, whereas the Lambertian signal decays exponentially with the observable geometric time delay, i.e.,}
{\color{black}
\begin{eqnarray}
C_{\rm spec}(\mathcal{R}, \Delta t_{\rm echo}) &=& w_s\,\mathcal{R}, \nonumber\\
C_{\rm Lamb}(\mathcal{R}, \Delta t_{\rm echo}) &\simeq& \frac{N K\,\mathcal{R}}{1-\mathcal{R}}\exp\!\left(-\kappa\,\frac{\Delta t_{\rm echo}}{t_g}\right),
\label{eq:15}
\end{eqnarray}
}
\textcolor{black}{where the Lambertian form holds for $\mathcal{R}\bar f\ll1-\mathcal{R}$ and $\epsilon\ll\epsilon_c$ (ensuring $\Gamma\to1$), with the additive constant $\mathcal{C}(a,\theta)$ of Eq.~(\ref{eq:6}) solely renormalizing its prefactor.} Equation~(\ref{eq:15}) dictates that a horizonless object is practically detectable only if $C$ exceeds the intrinsic dynamic range threshold $C_{\rm th}$ of the observing array\textcolor{black}{, which is fundamentally limited by thermal noise, atmospheric phase turbulence, and residual calibration errors}. Based on current and projected VLBI visibility amplitude calibrations \textcolor{black}{under these realistic conditions}, we adopt nominal contrast thresholds of $10^{-1}$ for the EHT, $3\times 10^{-2}$ for the ngEHT~\cite{Chael:2022meh,Roelofs:2022lux}, and $10^{-2}$ for the proposed space-based Black Hole Explorer (BHEX)~\cite{Johnson:2024ttr}.
\begin{table*}[t]
\color{black}
\caption{Prefactors of the unified contrast $C=w(a,i)\,\mathcal{R}(1-\eta_{\rm ret})/(1-\mathcal{R}\eta_{\rm ret})$: $w_s$ (specular, $\eta_{\rm ret}=0$) / $N$ (Lambertian, $\eta_{\rm ret}=1-\bar f$). $C$ is the mean brightness excess in the inner-shadow core (critical curve interior, more than $1\,GM/c^2$ from the curve) normalised by the mean $\mathcal{R}=0$ brightness of the photon-ring band ($|\Delta b|<0.5\,GM/c^2$). Averaged over $\epsilon\le10^{-3}$ (and $\mathcal{R}$); residual scatter $<1.5\%$. For $a=0.98$ the part of the flow inside the ergoregion also sends negative-energy photons to the membrane; they are neglected here ($\lesssim3\%$ of the membrane illumination).}
\label{tab:prefactors}
\begin{ruledtabular}
\begin{tabular}{cccccc}
$a$ & $i=17^\circ$ & $i=30^\circ$ & $i=40^\circ$ & $i=60^\circ$ & $i=75^\circ$ \\
\hline
0 & 0.616 / 0.815 & 0.565 / 0.717 & 0.527 / 0.635 & 0.482 / 0.508 & 0.475 / 0.458 \\
0.3 & 0.509 / 0.601 & 0.480 / 0.557 & 0.461 / 0.521 & 0.446 / 0.461 & 0.449 / 0.434 \\
0.5 & 0.421 / 0.436 & 0.410 / 0.433 & 0.403 / 0.427 & 0.399 / 0.409 & 0.404 / 0.397 \\
0.7 & 0.347 / 0.289 & 0.337 / 0.301 & 0.334 / 0.309 & 0.333 / 0.313 & 0.339 / 0.312 \\
0.9 & 0.358 / 0.169 & 0.306 / 0.170 & 0.268 / 0.160 & 0.224 / 0.130 & 0.204 / 0.111 \\
0.94 & 0.368 / 0.130 & 0.303 / 0.129 & 0.258 / 0.119 & 0.191 / 0.080 & 0.161 / 0.061 \\
0.98 & 0.378 / 0.065 & 0.298 / 0.065 & 0.241 / 0.056 & 0.156 / 0.028 & 0.116 / 0.018 \\
\end{tabular}
\end{ruledtabular}
\end{table*}

\par
\begin{figure*}[htbp]
\centering
\includegraphics[width=15cm,height=4.5cm]{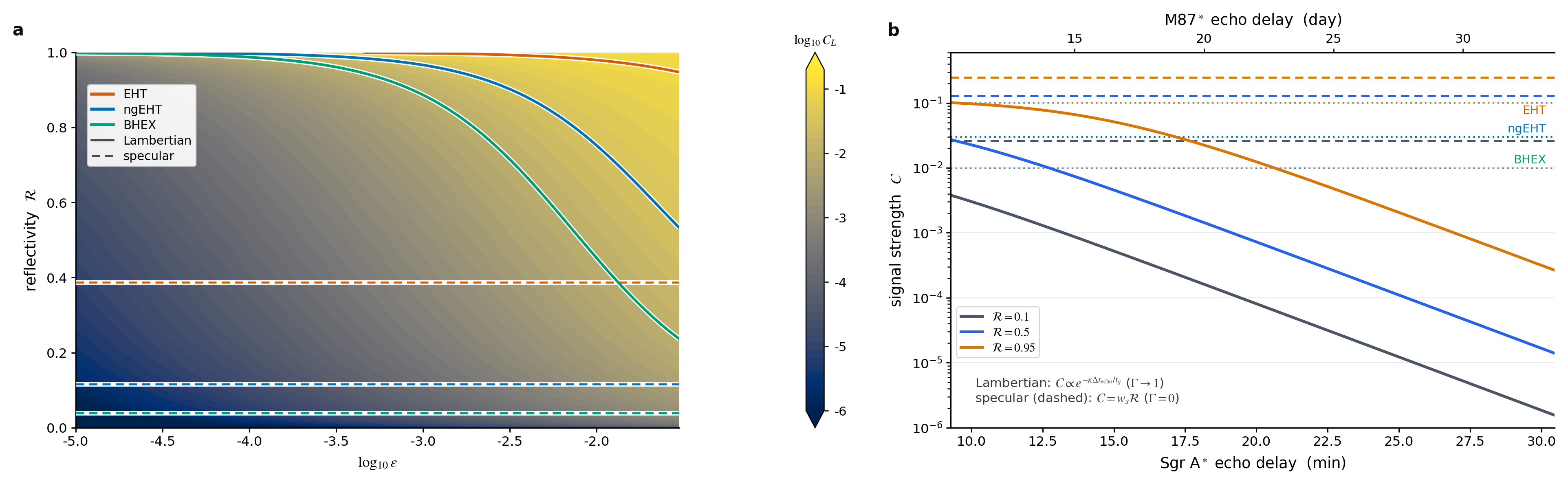}
\caption{The horizon discovery space quantifying the instrumental detectability \textcolor{black}{limits} of the quantum boundary\textcolor{black}{ ($a=0.94$, $i=40^\circ$)}. (a) Contours of the unified signal strength $C(\mathcal{R}, \epsilon)$, denoting the critical dynamic range limits requiblack to rule out horizonless models for the \textcolor{black}{current and next-generation VLBI arrays (}EHT, ngEHT, and BHEX\textcolor{black}{)}\textcolor{black}{: colour shows the Lambertian contrast, solid curves mark $C=C_{\rm th}$ (the region above each curve is detectable), and dashed horizontal lines mark the $\epsilon$-independent specular limit $\mathcal{R}=C_{\rm th}/w_s$}. \textcolor{black}{(b) Signal strength versus echo delay for $\mathcal{R}=0.1$, $0.5$ and $0.95$: Lambertian (solid), which approaches $C\propto\exp(-\kappa\Delta t_{\rm echo}/t_g)$ [Eq.~(\ref{eq:15})] at long delays, and specular (dashed), independent of $\Delta t_{\rm echo}$; dotted lines mark the array thresholds.} The extreme mass differential between the two targets dictates complementary constraints: static ultra-deep imaging probes the day-scale echoes of M87*, whereas dynamical time-domain monitoring targets the minute-scale temporal resonances of Sgr A*.}
\label{fig:8}
\end{figure*}

Figure~\ref{fig:8} visualizes this resulting horizon discovery space\textcolor{black}{ for $a=0.94$ and $i=40^\circ$, utilizing the calibrated $C(\mathcal{R},\epsilon)$ alongside the prefactors from Table~\ref{tab:prefactors}. The contour map (Fig.~\ref{fig:8}a) delineates the forecast detectability boundaries of the respective VLBI arrays. In the specular limit, the reach remains structurally independent of $\epsilon$: the minimum detectable reflectivities are $\mathcal{R}_{\min}=C_{\rm th}/w_s=0.39$, $0.12$, and $0.039$ for the EHT, ngEHT, and BHEX, respectively ($0.27$, $0.082$, and $0.027$ at M87$^\ast$'s inclination $i\simeq17^\circ$). In the Lambertian limit, the reach degrades exponentially toward the horizon: $\mathcal{R}_{\min}=0.98$, $0.75$, and $0.45$ at $\epsilon=10^{-2}$, and $0.998$, $0.97$, and $0.89$ at $\epsilon=10^{-3}$. Since $C\le N(a,i)\simeq0.12$ even for $\mathcal{R}\to1$, the core of a diffusely re-emitting membrane intrinsically acts as a low-contrast target. A membrane featuring partly specular and partly diffuse reflection properties naturally interpolates between these two limits.} Crucially, Fig.~\ref{fig:8}b translates this mathematical scaling into physical delay times for the two primary supermassive black hole targets: M87* ($t_g \approx 8.9$\,h) and Sgr A* ($t_g \approx 20.2$\,s)\textcolor{black}{; across $10^{-5}\le\epsilon\le3\times10^{-2}$, the respective echo delays span $10$--$34$ days and $9$--$30$ minutes}. This scaling inherently mandates a strict observational complementarity driven by their disparate masses. For M87*, theoretical echo delays manifest on the order of days to weeks. Because this timescale significantly exceeds the Earth-rotation aperture synthesis window ($\sim 1$\,day), the macroscopic echoes exist strictly in the steady-state regime. M87* is therefore the absolute optimal target for ultra-deep static imaging to measure the continuous inner shadow filling (as derived in Sec.~\ref{sec:3-1}). Conversely, the macroscopic echoes of Sgr A* unfold on minute timescales. While this extreme temporal compression effectively \textcolor{black}{blurs} static \textcolor{black}{interferometric} visibilities, it uniquely transforms Sgr A* into an ideal time-domain laboratory. For Sgr A*, a single continuous VLBI track \textcolor{black}{spans many echo delays}, enabling the extraction of $\mathcal{R}$ through dynamic visibility closures and light-curve periodicity\textcolor{black}{, provided that the deterministic periodic signals can be successfully isolated from the stochastic accretion black noise (a critical challenge we explicitly address in Section~\ref{sec:4})}.

\section{Time-Domain Constraints from Sgr A* Light Curves}
\label{sec:4}
\par
Alternative gravity theories and quantum-gravity models frequently pblackict exotic compact objects that lack an absolute mathematical horizon but possess a structublack quantum boundary at $r_b = r_+(1+\epsilon)$. \textcolor{black}{Crucially,} we demonstrate that the theoretical formalism connecting such boundaries to quantum decoherence is not merely a theoretical target for future spatial interferometry, but can be robustly translated into a time-domain signature \textcolor{black}{that is} immediately testable with existing high-cadence observations of Sgr A$^\ast$. In curved spacetime, if a quantum system is maintained outside a black hole, it continuously radiates soft (low-frequency) entangled fields across the Killing horizon. \textcolor{black}{Built upon} the rigorous field-theoretic framework developed by Ref.~\cite{Danielson:2022tdw}, this DSW decoherence rate is strictly determined by the two-point Wightman function of the near-horizon soft radiation~\cite{Danielson:2022sga}. The accumulated decoherence exponent $\mathcal{D}$ over a proper time $T$ is obtained by integrating the product of the thermal Unruh density of states $n(\omega)$ and the Fourier-transformed superposed source current $|\Delta \tilde{J}(\omega)|^2$:
\begin{equation}
\mathcal{D} = \int_0^\infty d\omega \, n(\omega) \big| \Delta \tilde{J}(\omega) \big|^2.
\label{eq:16}
\end{equation}
For a Kerr horizon, \textcolor{black}{the thermal factor $n(\omega)\simeq\kappa/(\pi\omega)$ at $\omega\ll\kappa$ makes the infrablack modes dominate}, leading to a secular, strictly linear decoherence growth $\mathcal{D} = \Gamma_{\rm dec}^{(0)} T$. Here, the intrinsic baseline decoherence rate scales \textcolor{black}{with the surface gravity, $\Gamma_{\rm dec}^{(0)} = C_0 \kappa$, where the prefactor $C_0$ depends on the charge (or mass) of the superposed body, its separation $d$, its position and, for $a\neq0$, on the spin~\cite{Danielson:2022tdw,Gralla:2023oya}; in the electromagnetic case the decoherence indeed vanishes in the extremal limit $\kappa\to0$~\cite{Gralla:2023oya}}.

\par
\begin{figure*}[htbp]
\centering
\includegraphics[width=15cm,height=5.5cm]{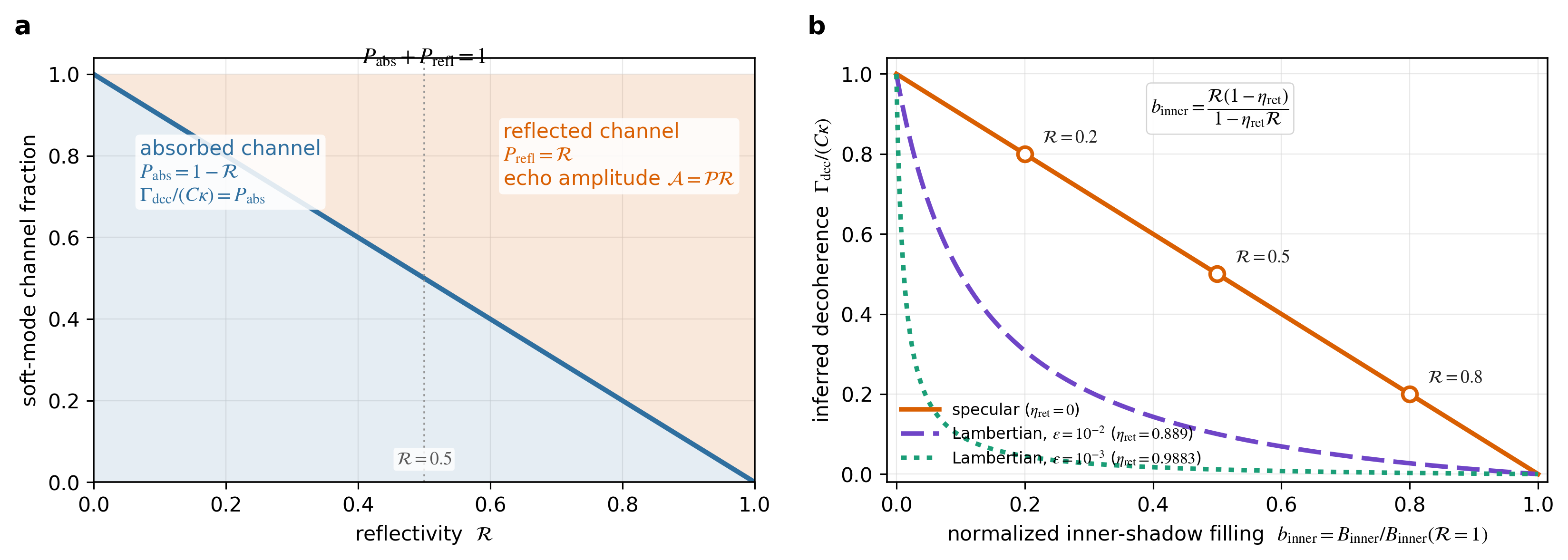}
\caption{The physical mapping between fundamental quantum decoherence and macroscopic observables governed by a single reflectivity parameter $\mathcal{R}$. (a) Soft-mode channel partitioning: the fraction absorbed by the boundary ($P_{\rm abs} = 1-\mathcal{R}$) acts as a causal sink driving irreversible quantum decoherence, while the reflected fraction ($P_{\rm refl} = \mathcal{R}$) preserves \textcolor{black}{quantum} information and generates classical \textcolor{black}{electromagnetic} echoes\textcolor{black}{ of amplitude $\mathcal{A}=\mathcal{P}\mathcal{R}$ [Eq.~(\ref{eq:19})]}. (b) The \textcolor{black}{invertible} dictionary: the strictly monotonic \textcolor{black}{relationship} $b_{\rm inner} = \mathcal{R}(1-\eta_{\rm ret}) / (1-\eta_{\rm ret}\mathcal{R})$ \textcolor{black}{[Eq.~(\ref{eq:10})], shown for a specular membrane ($\eta_{\rm ret}=0$; circles mark $\mathcal{R}=0.2$, $0.5$ and $0.8$) and for a Lambertian membrane with $a=0.94$ at $\epsilon=10^{-2}$ and $10^{-3}$ ($\eta_{\rm ret}=1-\bar f$),} allows observers to directly infer the microscopic information-loss rate $\Gamma_{\rm dec}/(C\kappa)$ from normalized spatial interferometric signatures\textcolor{black}{, once the reflection law and $\epsilon$ are fixed}.}
\label{fig:9}
\end{figure*}

To model a horizonless ECO, we must modify the classical absorber boundary condition. Consider a near-horizon membrane at $r_b$ characterized by a phenomenological reflectivity $\mathcal{R}(\omega)$. By energy and probability conservation, the incoming soft-mode channel is fundamentally partitioned into a transmitted (absorbed) fraction and a reflected fraction: $P_{\rm abs} + P_{\rm refl} \equiv (1-\mathcal{R}) + \mathcal{R} = 1$. The physical consequence for open quantum systems is that only the absorbed photons ($P_{\rm abs}$) permanently enter the unobservable causal sink to drive environmental decoherence~\cite{Fahn:2026rei}. Conversely, the reflected photons ($P_{\rm refl}$) return to the asymptotic exterior (future null infinity $\mathcal{I}^+$), preserving the quantum information within the observable universe while simultaneously generating macroscopic electromagnetic echoes (Fig.~\ref{fig:9}a). Applying this causal partition to the DSW integral, the modified decoherence exponent strictly becomes
\begin{equation}
\mathcal{D}_{\mathcal{R}} = \int_0^\infty d\omega \, \big(1 - \mathcal{R}(\omega)\big) n(\omega) \big| \Delta \tilde{J}(\omega) \big|^2.
\label{eq:17}
\end{equation}
\textcolor{black}{As physically justified in Section~\ref{sec:2},} operating within the dominant soft infrablack window ($\omega \lesssim \kappa$), we \textcolor{black}{safely} adopt the \textcolor{black}{macroscopic} gray-body approximation $\mathcal{R}(\omega) \simeq \mathcal{R}$. Factoring this \textcolor{black}{constant} out yields \textcolor{black}{our mapping dictionary, which inherits the phenomenological status of the channel partition discussed in Sec.~\ref{sec:2}}\textcolor{black}{:}
\begin{equation}
\Gamma_{\rm dec}(\mathcal{R}) = (1 - \mathcal{R}) C \kappa,
\label{eq:18}
\end{equation}
where $C$ incorporates purely kinematic coupling constants \textcolor{black}{($C=C_0$ for the gray-body boundary)}. Crucially, this microscopic dictionary maps directly onto macroscopic spatial imaging. \textcolor{black}{By Eq.~(\ref{eq:10}), the normalized inner shadow filling is $b_{\rm inner}\equiv B_{\rm inner}/B_{\rm inner}(\mathcal{R}=1) = \mathcal{R}(1-\eta_{\rm ret}) / (1-\eta_{\rm ret}\mathcal{R})$, with $\eta_{\rm ret}=0$ for a specular membrane and $\eta_{\rm ret}=1-\bar f(\epsilon;a)$ for a Lambertian one ($\eta_{\rm ret}=0.889$ and $0.988$ at $\epsilon=10^{-2}$ and $10^{-3}$ for $a=0.94$; Fig.~\ref{fig:9}b). Because $b_{\rm inner}$ is a strictly monotonic function of $\mathcal{R}$ at a fixed reflection law and $\epsilon$, the mapping is invertible. For a specular membrane it does not depend on $\epsilon$; for a diffuse one $\epsilon$ must be known, and it is provided independently by the echo delay measublack in the time domain (see below). Imaging and timing together therefore determine $\mathcal{R}$, and hence the microscopic decoherence capacity $\Gamma_{\rm dec}/(C\kappa)=1-\mathcal{R}$ of the boundary.}

\par
\begin{figure*}[htbp]
\centering
\includegraphics[width=15cm,height=10cm]{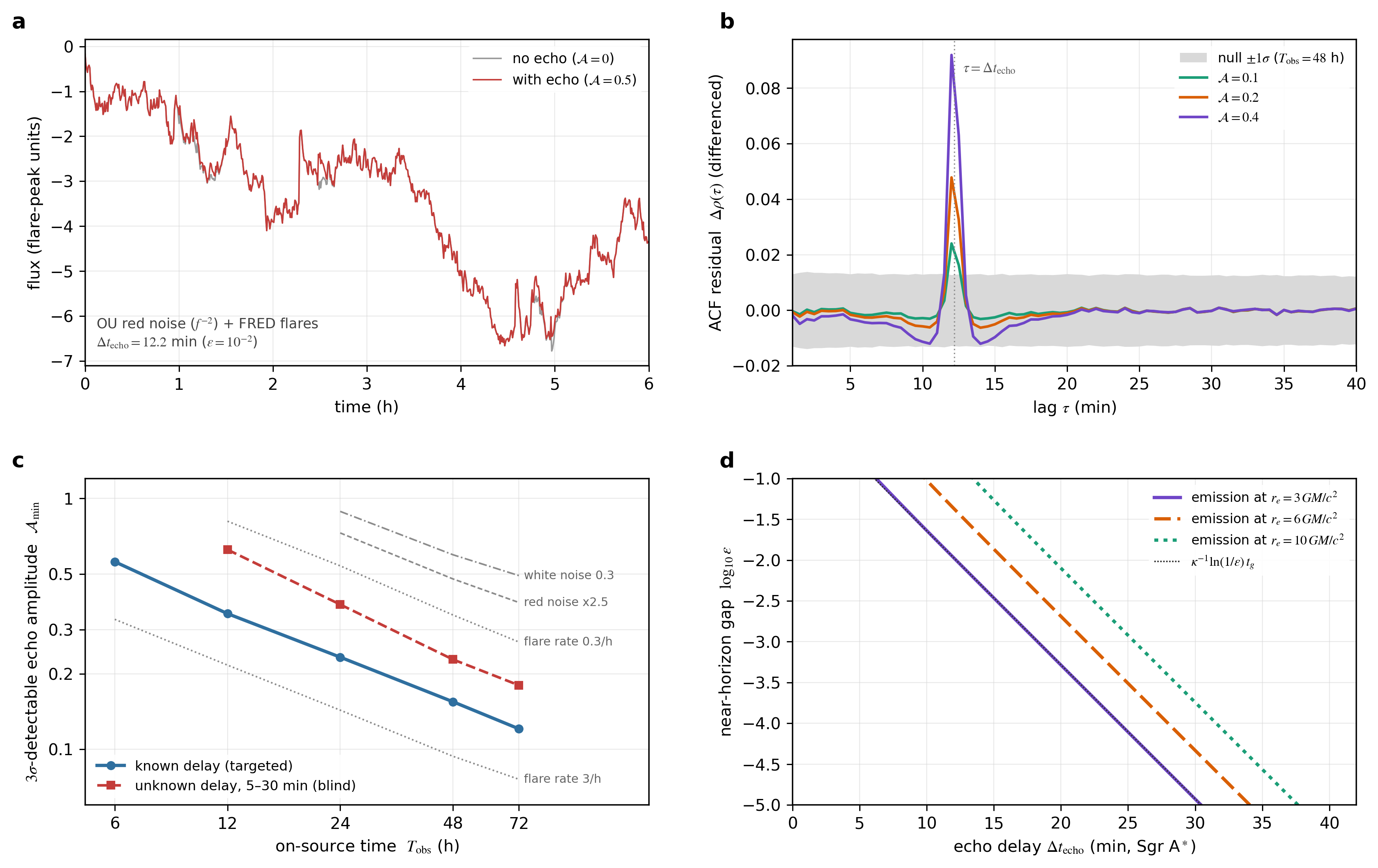}
\caption{The Sgr A$^\ast$ Time-Domain Detection Pipeline \textcolor{black}{demonstrating robustness against accretion black noise}. (a) A simulated \textcolor{black}{six-hour night of} intrinsic stochastic \textcolor{black}{flares} \textcolor{black}{(modeled as a FBLACK profile embedded in a $\sim f^{-2}$ black noise background)} \textcolor{black}{with single echoes ($\Delta t = 12.2$ min, $\mathcal{A} = 0.5$; grey: the same realization without echoes)}. (b) The ACF residual $\Delta\rho(\tau)$ \textcolor{black}{of the differenced light curves (mean of simulations with $T_{\rm obs}=48$\,h; grey band: $\pm1\sigma$ scatter of echo-free realizations) isolates the reflection echo as a resonant peak at $\tau = \Delta t_{\rm echo}$} (Eq.~\ref{eq:21}), \textcolor{black}{distinguishing} a reflective quantum membrane from an absorbing classical horizon \textcolor{black}{even in the presence of dominant black noise}. (c) Echo detectability \textcolor{black}{versus} continuous on-source time $T_{\rm obs}$\textcolor{black}{: minimum echo amplitude detected at $3\sigma$ with 50\% probability, for a known delay (targeted) and for a blind search over $5$--$30$\,min, with thresholds calibrated on $4\times10^4$ echo-free realizations per point; grey: targeted search for a lower or higher flare rate, stronger black noise and additional white noise}. (d) \textcolor{black}{Delay--gap relation used to invert the measublack lag, $\Delta t_{\rm echo}=\kappa^{-1}\ln(1/\epsilon)\,t_g+\mathcal{C}$, for principal null rays emitted at $r_e=3$, $6$ and $10\,GM/c^2$ ($a=0.94$); dotted: $\kappa^{-1}\ln(1/\epsilon)\,t_g$.}}
\label{fig:10}
\end{figure*}

While steady-state imaging constraints on M87$^\ast$ require ultra-high interferometric dynamic ranges, the reflection mechanism naturally yields an independent time-domain signature. If the boundary is partially reflective, stochastic hot-spot fluctuations or magnetic reconnection flares originating in the accretion flow \textcolor{black}{are partly intercepted by the membrane and returned toward the observer after the delay $\Delta t_{\rm echo}$. For a specular membrane every intercepted ray is reflected exactly once and escapes (Sec.~\ref{sec:3}), so each flare is followed by a single echo.} \textcolor{black}{Consequently,} the \textcolor{black}{observed} broad-band light curve $F(t)$ takes the \textcolor{black}{form}
\begin{equation}
\textcolor{black}{F(t) = L(t) + \mathcal{A}\,L(t - \Delta t_{\rm echo}), \qquad \mathcal{A} \equiv \mathcal{P}\mathcal{R},}
\label{eq:19}
\end{equation}
where $L(t)$ is the intrinsic primary flare profile, and \textcolor{black}{$\mathcal{A}$ is the echo amplitude: the reflectivity times the echo-to-direct flux ratio $\mathcal{P}$ that a perfectly reflecting membrane ($\mathcal{R}=1$) would produce. Full ray tracing of the flow of Sec.~\ref{sec:2} with $a=0.94$ and $\epsilon=10^{-2}$ gives $\mathcal{P}=0.17$, $0.13$ and $0.10$ for $i=17^\circ$, $40^\circ$ and $60^\circ$; for emission confined to $r\le3\,GM/c^2$, as expected for compact flares near the ISCO, $\mathcal{P}$ rises to $0.73$, $0.39$ and $0.19$, because a large fraction of the photons emitted there is directed toward the membrane. Diffuse re-emission does not produce a train of echoes with period $\Delta t_{\rm echo}$ either: re-emitted positive-energy photons that fall back onto the membrane do so after a median of $15$--$16\,t_g$ (90\% within $32\,t_g$) for $a=0.94$ and $\epsilon=10^{-2}$--$10^{-3}$, much less than $\Delta t_{\rm echo}=36$--$54\,t_g$, while negative-energy photons remain confined to the ergoregion; repeated re-processing therefore adds a smooth tail rather than discrete echoes. A multi-echo comb would require partial reflection of the waves at the angular-momentum barrier, which is exponentially small for $\omega M\gg1$, i.e., at millimeter and infrablack wavelengths~\cite{Cardoso:2019rvt}.} \textcolor{black}{In our simulation, the intrinsic primary flare $L(t)$ is modeled as a Fast Rise Exponential Decay (FBLACK) profile to mimic magnetic reconnection events, embedded in a stochastically generated black noise background with a power spectral density of $\sim f^{-2}$.} Figure~\ref{fig:10}a \textcolor{black}{shows one simulated six-hour night containing several flares and their echoes}. The observational feasibility of detecting this \textcolor{black}{delayed} signature is critically governed by the physical mass of the target. For Sgr A$^\ast$ (\textcolor{black}{$M \simeq 4.1 \times 10^6 M_\odot$}), the gravitational timescale is $t_g \simeq 20.2$\,s. Assuming a highly spinning background ($a=0.94$) and a characteristic quantum membrane gap $\epsilon = 10^{-2}$, the theoretical macroscopic echo delay mathematically evaluates to
\begin{equation}
\Delta t_{\rm echo} \simeq \frac{1}{\kappa} \ln\left(\frac{1}{\epsilon}\right) t_g \simeq 12.2 \text{ minutes}.
\label{eq:20}
\end{equation}
\textcolor{black}{From an observational standpoint,} this $\mathcal{O}(10)$ minute-scale temporal delay is phenomenally advantageous. It aligns perfectly with the typical near-infrablack and X-ray intrinsic flaring timescales of Sgr A$^\ast$, falling entirely within a single, continuous night-long observational track~\cite{Abuter:2018uum,Baganoff:2001kw}. \textcolor{black}{The additive constant $\mathcal{C}(a,\theta)$ of Eq.~(\ref{eq:6}) adds the light-travel time from the emission region: for principal null rays emitted at $r_e=3$, $6$ and $10\,GM/c^2$ the round-trip delay is $12.2$, $15.8$ and $19.4$\,min (Fig.~\ref{fig:10}d).}

\par
In realistic multi-wavelength observations, however, these discrete echoes are deeply buried within \textcolor{black}{the intrinsic} stochastic continuous black noise \textcolor{black}{exhibited by the accretion flow} (the typical $\sim f^{-2}$ \textcolor{black}{power spectral density}). \textcolor{black}{This severe black-noise background poses a fundamental challenge for direct time-domain identification.} \textcolor{black}{To rigorously break this degeneracy and} extract the boundary signature, \textcolor{black}{we mathematically compute} the Autocorrelation Function (ACF) of the continuous light curve, $\rho(\tau) = \langle F(t)F(t+\tau) \rangle$\textcolor{black}{, normalized to $\rho(0)=1$, after differencing the light curve, which whitens an $f^{-2}$ spectrum. Subtracting the mean ACF of echo-free realizations, $\langle\rho(\tau)\rangle_{\rm null}$, isolates the ACF residual $\Delta\rho(\tau)$, to which every flare contributes coherently at the same lag, yielding a resonant peak at the echo delay}, i.e.,
\begin{equation}
\textcolor{black}{\Delta\rho(\tau) \simeq \frac{\mathcal{A}f_{\rm fl}}{1+\mathcal{A}^2 f_{\rm fl}}\,\rho_L(\tau - \Delta t_{\rm echo}), \qquad \tau>0,}
\label{eq:21}
\end{equation}
where \textcolor{black}{$\rho_L$ is the normalized autocorrelation of the (differenced) flare profile and $f_{\rm fl}$ is the fraction of the (differenced) variance contributed by flares ($f_{\rm fl}\simeq0.36$ in our fiducial model)}. While stochastic accretion \textcolor{black}{black} noise elevates the \textcolor{black}{broadband baseline and heavily biases the} zero-lag ($\tau=0$) peak, it intrinsically lacks \textcolor{black}{strict} temporal coherence\textcolor{black}{. Therefore, it} cannot \textcolor{black}{mathematically} replicate the \textcolor{black}{sharp resonance} \textcolor{black}{that strictly emerges} \textcolor{black}{at the same lag $\tau=\Delta t_{\rm echo}$ for every flare}\textcolor{black}{, rendering the ACF residual highly resilient against stochastic astrophysical backgrounds}.

\par
Figure~\ref{fig:10} presents a comprehensive time-domain signal-extraction pipeline applied to synthetic Sgr A$^\ast$ light curves. \textcolor{black}{The light curves are sampled every $0.5$\,min during continuous six-hour nights and consist of an Ornstein--Uhlenbeck black noise (power spectrum $\propto f^{-2}$ above the break at $1/(2\pi\times4\,{\rm h})$, comparable to the coherence timescale of a few hours found for the near-infrablack variability of Sgr A$^\ast$~\cite{Witzel:2018kzq}, with an rms of twice the median flare peak), Poisson-distributed FBLACK flares (rate $1\,{\rm h}^{-1}$, rise and decay times of $0.5$ and $2.5$\,min, log-normal peak amplitudes with $\sigma_{\ln}=0.5$), white noise of $0.05$ median flare peaks per sample, and the echo of Eq.~(\ref{eq:19}).} The derived ACF residual cleanly isolates the \textcolor{black}{echo peak} (Fig.~\ref{fig:10}b). The statistical detectability, quantified by the primary echo-peak Signal-to-Noise Ratio (SNR), \textcolor{black}{improves with} the effective continuous integration time $T_{\rm obs}$\textcolor{black}{, effectively suppressing the uncorrelated black noise over long epochs}. \textcolor{black}{Because the null distribution is non-Gaussian, we calibrate the $3\sigma$ threshold (one-sided false-alarm probability $1.35\times10^{-3}$) empirically from $4\times10^4$ echo-free realizations for each $T_{\rm obs}$. As shown in Fig.~\ref{fig:10}c, an echo is detected with 50\% probability for $\mathcal{A}\ge0.35$, $0.23$ and $0.12$ after $T_{\rm obs}=12$, $24$ and $72$\,h if the delay is known (targeted search), and for $\mathcal{A}\ge0.63$, $0.38$ and $0.18$ if the delay is scanned over $5$--$30$\,min (blind search, including the look-elsewhere effect), approximately as $\mathcal{A}_{\min}\propto T_{\rm obs}^{-0.6}$ (targeted) and $T_{\rm obs}^{-0.7}$ (blind). An echo amplitude $\mathcal{A}=0.3$ requires $\simeq15$\,h (targeted) or $\simeq33$\,h (blind) of on-source time. Since $\mathcal{A}=\mathcal{P}\mathcal{R}$, this corresponds to $\mathcal{R}\gtrsim0.8$ for compact flares at $i=40^\circ$ ($\mathcal{P}=0.39$), and 72\,h of targeted monitoring reach $\mathcal{R}\simeq0.3$ for such flares but only $\mathcal{R}\simeq0.9$ for emission spread over the whole flow ($\mathcal{P}=0.13$).} Furthermore, \textcolor{black}{robustly} \textcolor{black}{locating the ACF peak measures $\Delta t_{\rm echo}$, which Eq.~(\ref{eq:6}) converts into the near-horizon gap $\epsilon$ (Fig.~\ref{fig:10}d). Because the additive constant $\mathcal{C}(a,\theta)$ depends on where the flare is emitted, the inversion requires the emission radius: for $a=0.94$, an uncertainty $3\le r_e\le10\,GM/c^2$ changes the inferblack $\epsilon$ by a factor of $\simeq15$.} Consequently, an empirical search for horizon-scale quantum structure can be \textcolor{black}{carried out} \textit{today} by applying this ACF pipeline to existing multi-epoch, high-cadence light curves from facilities such as GRAVITY (NIR) and the EHT (sub-mm)~\cite{Abuter:2018uum,Roelofs:2022lux}. \textcolor{black}{It should be noted that our synthetic SNR threshold primarily accounts for the dominant intrinsic accretion black noise. In practical observational campaigns, residual instrumental white noise and atmospheric phase fluctuations may slightly degrade the ACF temporal coherence, potentially necessitating joint multi-epoch observations to rigorously secure the $3\sigma$ statistical detection threshold.} \textcolor{black}{Our simulations quantify these effects (Fig.~\ref{fig:10}c): white noise of $0.3$ median flare peaks per sample raises $\mathcal{A}_{\min}$ by a factor of $\simeq4$, a $2.5$ times stronger black noise by $\simeq3$ and a flare rate of $0.3\,{\rm h}^{-1}$ by $\simeq2$, whereas a rate of $3\,{\rm h}^{-1}$ lowers it by $\simeq1.6$. The thresholds therefore depend on the flare statistics and must be recalibrated for a given data set, e.g., with injection--recovery tests.}

\section{Conclusion}
\label{sec:5}
\par
The \textcolor{black}{true} nature of the black hole event horizon remains one of the most profound open questions in fundamental physics. While current interferometric observations have definitively confirmed the existence of the \textcolor{black}{macroscopic shadow and the surrounding emission ring}, the geometric shadow alone cannot \textcolor{black}{unambiguously} differentiate a true information-absorbing horizon from an ultra-compact reflective boundary. In this paper, we \textcolor{black}{have} constructed a \textcolor{black}{rigorous}, observationally \textcolor{black}{verifiable}\textcolor{black}{, phenomenological} dictionary connecting the microscopic quantum decoherence of the horizon to macroscopic electromagnetic observables.

\par
By introducing a unified phenomenological reflectivity parameter $\mathcal{R}$, we demonstrated that the capacity of a black hole to act as an irreversible causal sink (driving the DSW decoherence) is strictly anti-correlated with the generation of delayed geometric echoes. We embedded this boundary condition into a highly adaptive general relativistic ray-tracing framework in the Kerr spacetime\textcolor{black}{, in which the null condition is enforced at every reflection, and bracketed the unknown microscopic reflection law by two limits, specular reflection and diffuse (Lambertian) re-emission}. Our analysis yields three principal conclusions\textcolor{black}{:} First, in the steady-state spatial domain, \textcolor{black}{the light returned by the membrane elevates the flux within the inner shadow: linearly in $\mathcal{R}$ for specular reflection, for which every captublack ray is reflected exactly once and escapes, and through a geometric series with return fraction $\eta_{\rm ret}=1-\bar f$ for diffuse re-emission [Eq.~(\ref{eq:10})]. For $a=0.94$, $i=40^\circ$ and $\mathcal{R}=0.8$, the flux within $3\,GM/c^2$ of the image centre rises by factors of $1.94$ (specular) and $1.48$ (Lambertian).} Crucially, \textcolor{black}{as long as the membrane lies inside the photon shell,} this macroscopic inner-shadow filling \textcolor{black}{leaves every pixel outside the critical curve unchanged, so that the photon subrings, which are governed exclusively by the background metric, provide a boundary-independent reference}. In the Fourier domain, this allows \textcolor{black}{next-generation} VLBI arrays to utilize ultra-long baselines to calibrate the background Kerr geometry\textcolor{black}{ through the photon-ring diameter and ringing period, which do not depend on $\mathcal{R}$}, while independently extracting the boundary parameter $\mathcal{R}$ from the low-frequency visibility envelope. \textcolor{black}{Furthermore, this low-baseline signature is observationally accessible: on Earth-diameter baselines to M87$^\ast$ ($u\lesssim0.27$ cycles$/GMc^{-2}$ at 345~GHz), a membrane with $\mathcal{R}=0.8$ changes the normalized visibility amplitude by $\simeq11\%$ (specular) or $\simeq17\%$ (Lambertian), so that a $3\sigma$ detection on a single visibility requires an SNR of $\simeq28$ or $\simeq17$, well within reach of dense $(u,v)$-coverage arrays. The main systematic is the station-gain calibration, whereas accretion non-stationarity, interstellar scattering and plasma refraction are expected to affect this signature only weakly.}

\par
Second, we defined a unified parameter space to evaluate the instrumental detectability of horizon-scale quantum structures \textcolor{black}{under realistic observational constraints}. \textcolor{black}{The effective exponent $\Gamma=\partial\ln C/\partial\ln\epsilon$ of the signal strength is not a universal number but is fixed by the reflection law through Liouville's theorem: $\Gamma=0$ for specular reflection, because the capture and escape cones shrink together and the blackshift on the way in is exactly undone on the way out, whereas $\Gamma\to1$ for diffuse re-emission, because the \'etendue of the escape cone scales as the squablack lapse, $\propto\epsilon$. We provided a fitting formula and tables for $\Gamma(a,i)$ over $0\le a\le0.98$ and $17^\circ\le i\le75^\circ$ [Eq.~(\ref{eq:13c}) and Tables~\ref{tab:gamma_fit}--\ref{tab:prefactors}]. For a specular membrane the minimum detectable reflectivity does not depend on $\epsilon$, $\mathcal{R}_{\min}=0.39$, $0.12$ and $0.039$ for the EHT, ngEHT and BHEX thresholds ($a=0.94$, $i=40^\circ$), whereas for a diffuse membrane it degrades exponentially with the echo delay, $C\propto\exp(-\kappa\Delta t_{\rm echo}/t_g)$, and already reaches $\mathcal{R}_{\min}=0.98$, $0.75$ and $0.45$ at $\epsilon=10^{-2}$.} We found that static imaging constraints require \textcolor{black}{contrast sensitivities of $10^{-2}$--$10^{-1}$}, placing \textcolor{black}{stringent} \textcolor{black}{requirements on the baseline sensitivity and calibration fidelity for} the upcoming ngEHT and the proposed space-based BHEX missions. Finally, the extreme mass differential between M87$^\ast$ and Sgr A$^\ast$ dictates a fundamental observational complementarity\textcolor{black}{: for $a=0.94$ and $10^{-5}\le\epsilon\le3\times10^{-2}$ the echo delays span $10$--$34$ days and $9$--$30$ minutes, respectively}. While the day-scale echo delays of M87$^\ast$ are optimal for static spatial interferometry, the minute-scale temporal delays of Sgr A$^\ast$ transform it into an ideal time-domain laboratory. By simulating realistic stochastic accretion flares \textcolor{black}{deeply embedded within severe black-noise backgrounds}, we demonstrated that the \textcolor{black}{delayed echo of each flare, which for a specular membrane is a single echo of amplitude $\mathcal{A}=\mathcal{P}\mathcal{R}$ rather than a periodic comb,} can be robustly isolated using \textcolor{black}{our tailoblack} ACF residual pipeline. Our statistical analysis\textcolor{black}{, with $3\sigma$ thresholds calibrated on $4\times10^4$ echo-free light curves, shows that an echo amplitude $\mathcal{A}\gtrsim0.3$ is detected (with 50\% probability) after $\simeq15$ hours of continuous on-source monitoring if the delay is known and after $\simeq33$ hours in a blind search, and that $72$ hours reach $\mathcal{A}\simeq0.12$, i.e., $\mathcal{R}\simeq0.3$ for compact flares near the ISCO ($a=0.94$, $i=40^\circ$)}\textcolor{black}{, proving the method's high resilience against uncorrelated astrophysical variations. Furthermore, in practical ground-based observational scenarios, uncalibrated instrumental white noise and atmospheric phase fluctuations can be effectively mitigated through joint multi-epoch campaigns to secure this theoretical limit}. \textcolor{black}{These thresholds depend on the flare rate and on the noise levels and must be recalibrated for a given data set with injection--recovery tests, and converting the measublack delay into $\epsilon$ further requires the emission radius of the flares.}

\par
Ultimately, the theoretical framework presented here bridges the gap between quantum information theory and observational astrophysics. Applying the proposed ACF pipeline to existing and forthcoming multi-wavelength, high-cadence light curves (e.g., from GRAVITY and the EHT)\textcolor{black}{, together with deep static imaging of the inner shadow of M87$^\ast$,} provides an immediate, empirical pathway to \textcolor{black}{definitively} constrain the \textcolor{black}{quantum} existence of the event horizon. \textcolor{black}{Our forecasts rest on stationary, optically thin emission models and on geometric optics at the membrane; extending them to general-relativistic magnetohydrodynamic flows and to wave-optics reflection at long wavelengths is left for future work.}

\begin{acknowledgments}
This work is supported by the National Natural Science Foundation of China (Grant No. 12505060, 12375043, 12575069, 12175027, 11875010), the Fund Project of Chongqing Normal University (Grant Number: 24XLB033, 26XLB001), Chongqing Natural Science Foundation General Program (Grant No. CSTB2025NSCQ-GPX1019, and Science and Technology Research Project of Chongqing Municipal Education Commission (Grant Number: KJQN202500563).
\end{acknowledgments}



\nocite{*}

\bibliography{apssamp}



\end{document}